\documentclass{iopart}
\usepackage{iopams}
\usepackage{epsf}
\usepackage{euscript}
\usepackage{graphicx}

\newcommand{\td}{\tilde{\tau}_w}
\newcommand{\Pt}{{\mathcal P}_{w}(\tilde{\tau}_w)}
\begin{document}
\renewcommand{\baselinestretch}{1.5}
\review[Statistics of Resonances and Delay Times in Random Media: Beyond Random Matrix Theory]
{Statistics of Resonances and Delay Times in Random Media: Beyond Random Matrix Theory}
\author{Tsampikos Kottos} 
\address{
Department of Physics, Wesleyan University, Middletown, Connecticut 06459-0155, USA and\\
Max-Planck-Institute for Dynamics and Self-Organization, Bunsenstrasse 10, D-37073 G\"ottingen,
Germany}
\date{\today }

\begin{abstract}
We review recent developments on quantum scattering from mesoscopic systems. Various spatial
geometries whose closed analogs shows diffusive, localized or critical behavior are 
considered. These are features that cannot be described by the universal Random Matrix 
Theory results. Instead one has to go beyond this approximation and incorporate them in
a non-perturbative way. Here, we pay particular emphasis to the traces of these non-universal 
characteristics, in the distribution of the Wigner delay times and resonance widths. The 
former quantity captures time dependent aspects of quantum scattering while the latter is 
associated with the poles of the scattering matrix.

\hspace {-0.5cm} submitted to J. Phys. A Special Issue -- Aspects of Quantum Chaotic Scattering
\end{abstract}

%=======================================================================================
\section{\bf Introduction}
\label{sec:introduction}

Quantum mechanical scattering in systems with complex internal dynamics has been a subject of
intensive research activity for a number of years. This interest was motivated by various areas
of physics, ranging from nuclear \cite{MW69}, atomic \cite{atomic} and molecular \cite{molecular}
physics, to mesoscopics \cite{B97}, quantum chaos \cite{S89,KS00}, and classical wave scattering
\cite{S99,DS90}. Recently, the interest in this subject was renewed due to technological developments
in quantum optics associated with the construction of new type of lasers \cite{NS97,WAL95} and
experimental investigation of atoms in optical lattices \cite{Raizen}.

The most fundamental object characterizing the process of quantum scattering is the unitary $S$
-matrix relating the amplitudes of incoming waves to the amplitudes of outgoing waves. The recent
advances in mesoscopic physics and quantum chaos, led to a fast development of powerful theoretical
techniques which allow us to understand the statistical properties of the $S-$matrix. At present, 
there are two complementary theoretical tools employed to calculate statistical properties of the 
$S-$matrix, namely the semiclassical and the stochastic approach. The starting point of the first 
is a representation of the $S-$ matrix elements in terms of a sum of classical orbits \cite{S89,KS00} 
while the latter exploits the similarity with ensembles of Random Matrices \cite{FS97}. At the 
same time, recent experimental progress allowed a direct comparison of the theoretical predictions
with actual experimental results. Microwave experiments (see \cite{KSW05,CG05} and references therein) 
offer a unique possibility to check even details of the existing theories in cases, where this is 
hardly possible by other methods while they raise new challenging questions (see for example \cite{KSW05,CG05,HZAOA05}). 

Apart from the study of the distribution of the $S-$matrix elements, resonance widths and Wigner
delay times distributions gained also a lot of attention. The latter quantity captures the time-
dependent aspects of quantum scattering. It can be interpreted as the typical time an almost 
monochromatic wave packet remains in the interaction region. Resonances are defined as poles of 
the $S$-matrix occurring at complex energies ${\cal E}_n = E_n - \frac i2 \Gamma_n$, where $E_n$ 
is the position and $\Gamma_n$ the width of the resonance. They correspond to "eigenstates" of 
the open system that decay in time due to the coupling to the "outside world" and they are related 
to conductance fluctuations and current relaxation \cite{AKL91}. For chaotic/ballistic systems 
Random Matrix Theory (RMT) is applicable, and the distributions of resonance widths ${\cal P}
(\Gamma)$ and Wigner delay times ${\cal P}(\tau)$ are known. A review of the RMT results can be 
found in \cite{FS97} (see also \cite{KS00}). 

In this contributions we aim in giving an overview of the recent developments in scattering
from open samples in conditions where RMT is not applicable and deviations from "universality" 
due to the appearance of localization is apparent. Although our presentation is focus on random 
media, one has to keep always in mind that these results valid also for dynamical systems with 
chaotic classical limit. Despite the fact that these systems are deterministic (in contrast 
to random media where randomness is "builded" up with the system) localization occurs due to 
complicated interference effects created by the underlying classical chaotic dynamics and for 
this reason is termed {\it dynamical localization} \cite{I90}. 

The observables that will be in the focus of our presentation are the distributions of resonance 
widths and delay times. We 
consider various spatial geometries and models whose closed analogs show features such as 
diffusion, criticality or localization. A short overview of localization theory and the 
definitions of the various regimes is given in Section \ref{sec:loctheory}. In the next Section 
\ref{sec:models} we present the mathematical formalism associated with the scattering process, 
and define the quantities of interest. In Section \ref{sec:resonance} we review the consequences 
of localization in the resonance width distribution. The corresponding results for the delay 
times are analyzed in \ref{sec:delay}. Finally in Section \ref{sec:quasi} we present some 
results for quasi-periodic systems at criticality. Our conclusions are given at the last Section 
\ref{sec:conclusions}

%=======================================================================================
\section{\bf Various Regimes in Localization Theory: A brief overview}
\label{sec:loctheory}

Localization of waves has always been among the most difficult yet most fascinating topics
in the study of wave propagation in disordered media. The first studies dealt with infinite
media, showing that localization is always achieved in one and two dimensions but that a 
minimum amount of disorder is required in dimensions larger than two \cite{A58,KM93,AALR79}. 
Its main feature is that the eigenfunctions of a disordered medium in the localized regime, 
are characterized by an exponential decay in space i.e. $|\Psi_n({\bf r})|\sim \exp(-|{\bf r}
-{\bf r_0}|/\xi)$, where $\xi$ is the localization length. A direct consequence
is that transmission is inhibited and the system behaves as an insulator. 

The fingerprints of localization in various quantities associated with the closed systems have 
been identified and quite well understood. Detailed numerical and theoretical studies gave a 
clear picture how the statistical properties of these quantities change as the disorder strength 
increase (for a recent review see \cite{M00} and references therein). For weak disorder, such 
that the mean 
free path $l_{\rm mean}$ is larger than any physical length scale, the system is in the {\it 
ballistic regime}. An equivalent definition of this regime is given through the condition $g= 
E_{\rm Th}/\Delta\rightarrow \infty$. The dimensionless ratio $g$ is known as the Thouless 
conductance, $\Delta$ is the mean level spacing and $E_{\rm Th}$ is the co-called Thouless energy. 
The latter plays a prominent role in the theory of random media, and is inverse proportional 
to the time needed for an excitation to propagate through the entire sample. In this regime
the predictions of Random Matrix Theory (RMT) were shown to describe accurately the statistical 
properties of various observables. For example, the eigenstates are extended all over the 
system, the eigenvalue spacing distribution follow with a good accuracy the famous Wigner 
surmise \cite{P65,E83} etc.

As the disorder increases the system becomes {\it diffusive}. This regime is characterized
by the condition that the system size $L$ is larger than the mean free path $l_{\rm mean}$ 
but still smaller than the localization length $\xi$ i.e. $l_{\rm mean}\ll L \ll \xi$. Using 
the powerful sigma-model approach it was explained how the deviations from 
the RMT results raise \cite{M00,FE95}. Detailed numerical experiments \cite{OKG02,KOG03,UMRS00}
verified the theoretical predictions. These deviations are specifically strong at the far 
``tails'' of the distribution of the eigenfunction intensities as well as of some related 
quantities and are signatures of the underlying classical diffusive dynamics \cite{OKG02,
KOG03,UMRS00}. They were shown to be related with anomalously localized states termed {\it 
pre-localized} states. In \cite{OKG02,KOG03} it was found that pre-localized states are also 
present in quantum systems with deterministic chaotic dynamics. As far as the spectral 
correlations are concerned it has been shown that there are large deviations above the 
Thouless energy $E_{\rm Th}=\hbar/\tau_D$ where $\tau_{\rm Th}=\hbar D/L^2$ is time to diffuse 
through the system with diffusion constant $D$ \cite{E83,AS86}. The Thouless conductance
$g$ is related with the latter as $g= D L^{d-2}$ where $d$ is the dimensionality of the system.
Rapid development in microwave techniques allowed for a direct observation of some of these 
predictions, in microwave experiments \cite{SS92,AGHHLRSW95,KKS95,PS00}. 

The deviations of the level and eigenfunction statistics from their RMT form, strengthen with 
increasing disorder and become especially pronounced at the {\it localization} regime . In 
this regime, inhibition of wave diffusion due to interference of multiple scattering waves 
take place \cite{A58}. The resulting scenario depends strongly on the dimensionality of the sample 
and the disorder strength. It turns out that the eigenstates are exponentially localized in 
low dimensional systems even for arbitrary weak disorder. As a matter of fact the localization
regime is defined through the condition that $L< \xi$. One of the consequences following 
from this fact is the prediction that the conductance of a sample goes exponentially to zero 
with increasing its length and the sample behaves as an insulator. In contrast, disordered 
single-particle systems in more than two-dimensions exhibits a reacher behavior. If the 
disorder is weak enough there is no localization and the system has a metallic behavior while
for strong disorder strength we recover the localization regime. The transition point from
a metallic to localized behavior is of special interest and is called Metal-Insulator Transition 
(MIT). 

The MIT where the phase transition from localized to extended states occurs, is characterized
by remarkably rich critical properties. In particular, 
the level statistics acquires a scale-independent form with distinct critical features 
\cite{M00,AS86,CKL96} while the eigenfunctions show strong fluctuations on all length scales 
and represent multi-fractal distributions \cite{M00,FE95,W80,SG91,EM00}. As a matter of fact,
a connection between multi-fractality and statistical properties of eigenvalues at MIT has
been recently established \cite{CKL96}.
The multi-fractal structure of the eigenfunctions is usually quantified by studying the size 
dependence of the so-called participation numbers (PN)
\begin{equation}
\label{PN}
{\cal N}_q = \left(\int \left|\psi({\bf r})\right|^{2q} d{\bf r}\right)^{-1}
\propto L^{(q-1)D_{q}}
\end{equation}
where $L$ is the linear size of the system and $D_q$ are the multi-fractal dimensions of the
eigenfunction $\psi({\bf r})$. Among all the dimensions, the correlation dimension
$D_2$ plays the most prominent role. The corresponding PN is roughly equal to the number of
non-zero eigenfunction components, and therefore is a good and widely accepted measure of the
extension of the states. At the same time, $D_2$ manifest itself in a variety of other physical
observables. As examples we mention the statistical properties of the spectrum \cite{M00,CKL96}, 
and the anomalous spreading of a wave-packet, and the spatial dispersion of the diffusion 
coefficient \cite{HK99}.

At the same time a considerable effort was made to understand the 
shape of the conductance distribution ${\cal P}(g)$ at MIT \cite{AKL91,BHMM01,RMS01}. However, 
it is still unclear whether the limiting ${\cal P}(g)$ is entirely universal, i.e. dependent 
only on the dimensionality and symmetry class, as required by the one-parameter scaling
theory of localization \cite{A58}. The latter is one of the major achievements
in the long history of studying the MIT. Its basic assumption is that close to
the MIT the change of the conductance $g$ with the sample size $L$ depends only
on the conductance itself, and not separately on energy, disorder, size and
shape of the sample, the mean free path etc.

Although a lot of studies have been devoted to the analysis of eigenfunctions and eigenvalues 
and of conductance, the properties of resonances, and delay times were left unexplored until 
recently. Nevertheless it was clear from the very beginning that their statistical properties 
depend strongly on the nature of the states of the finite system "in isolation". Thus Anderson 
localization must leave its fingerprints in these quantities which after all reflect the 
"leakage" of the waves to the leads, through the sample boundaries. In the next sections we 
will review the outcome of these studies and their deviations from the RMT predictions due to 
non-universal features. 

%=======================================================================================
\section{\bf Quantum Scattering: Basic Concepts }
\label{sec:models}

The scattering $S-$matrix relates the outgoing wave amplitudes to the incoming wave amplitudes. 
Assuming $M$ open channels, one can show that the $M\times M$ scattering matrix can be written 
in the form \cite{MW69}
\begin{equation}
\label{smatrix}
S({\cal E})= {\bf 1} - 2i\pi V^{\dagger} {1\over {\cal E}-H_{\rm eff}} V; \quad 
H_{\rm eff}=H_0 - i\pi VV^{\dagger}
\end{equation}
Here, $H_0$ stands for an $N-$dimensional self-adjoint Hamiltonian describing the closed 
counterpart of the system under consideration, ${\cal E}$ stands for the energy of the 
incoming waves, and $V$ is an $M \times N$ operator that contains matrix elements coupling 
the internal motion to one of the open $M$ channels. In principle, the matrix elements of 
the operator $V$ depends on energy. However, since this dependence is very weak (far away 
from channel thresholds), we can ignore it and consider $V_{i,j}$ to be energy-independent.
${\bf 1}$ is the $M\times M$ unit matrix. For a detail derivation of the scattering matrix 
for the case of a tight-binding model see \cite{SR03} while for a chaotic cavity see \cite{S03}. 
It is easy to verify that form (\ref{smatrix}) ensures the unitarity of the scattering matrix 
i.e. $S^{\dagger}S={\bf 1}$, provided the energy ${\cal E}$ takes only real values. When 
one allows the energy parameter to have a nonzero imaginary part, the $S-$matrix unitarity 
is immediately lost. Having ${\cal I}m {\cal E}>0$ correspond to the physical situation of 
uniform dumping inside the system \cite{DS90,SFS05} and it is responsible for the losses of 
the outgoing flux of the particles as compared to the incoming flux. The "dual" case 
${\cal I}m {\cal E}<0$ correspond to uniform amplification. The balance between the two fluxes 
is precisely the physical mechanism behind the $S-$matrix unitarity.

The poles of the scattering matrix $S$ are associated with the formation of resonance states.
They represent long-lived intermediate states to which bound states of a closed system are 
converted due to coupling to continua. Due to causality, they are located in the lower
half plane i.e. ${\cal E}_n = E_n - \frac i2 \Gamma_n$, where $E_n$ and $\Gamma_n$ are the 
position and the width of the resonances, respectively. They are solutions of the following 
secular equation
\begin{equation}
\label{poleseq}
\det ({\cal E}-H_{\rm eff}({\cal E}))=0.
\end{equation}
where the resonance width $\Gamma$ is inverse proportional to the lifetime of the corresponding 
resonance state. From Eqs.~(\ref{smatrix},\ref{poleseq}) it is clear that the formation of 
resonances is closely related to the internal dynamics inside the scattering region which 
is governed by $H_0$.

Another quantity that will be in the focus of this contribution is the Wigner delay time 
\cite{W55} and its variations (for an overview on the various definitions of delay times 
and their physical importance see \cite{CN02}). It captures the time-dependent aspects of 
quantum scattering. It can be interpreted as the typical time an {\it almost monochromatic} 
wave packet remains in the interaction region.  Formally the Wigner delay time $\tau_W$ is 
related with the energetic derivative of the total phase of the unimodular $S-$matrix
\begin{equation}
\label{Wtau}
\tau_W(E)= {1\over M} {\rm Tr} Q(E);\quad Q(E)=-i\hbar S^{\dagger}(E) {dS(E)\over dE}
\end{equation}
where $Q(E)$ is called the Wigner-Smith time delay matrix \cite{W55}. Its eigenvalues 
$\tau_q$, are called
proper delay times, and correspond to the time the particle dwells at a particular channel $q=1,
\cdots,M$. Alternatively, one can also define the {\it partial} delay times $\tau_q^{\rm p}$
as the energetic derivatives of the eigenphases $\{\theta_q\}, q=1,\cdots,M$ of the unimodular 
$S-$matrix i.e. $\tau_q^{\rm p} = \partial \theta_q /\partial E$ \cite{FS97}. Beyond the
one-channel case, proper and partial delay times differ, although the sum of partial/proper
delay times over all $M$ scattering channels are always equal and yield the Wigner delay time.

%=======================================================================================
\section{\bf Resonances}
\label{sec:resonance}

In this section we analyze the distribution of resonances ${\cal P}(\Gamma)$. The properties 
of resonances are of fundamental as well as technological interest. One can show that they 
determine the conductance fluctuations of a quantum dot in the Coulomb blockade regime 
\cite{ABG02}, or the current relaxation. The latter study constitute a fundamental source 
of physical information for systems which are coupled to a continuum via metallic leads or 
absorbing boundaries. While the radioactive decay is a prominent paradigm, more recent examples 
include atoms in optically generated lattices and billiards \cite{RSN97,F01}, the ionization 
of molecular Rydberg states \cite{B91}, photoluminescence spectroscopy of excitation relaxation 
in semiconductor quantum dots and wires \cite{B99}, and pulse propagation studies with 
electromagnetic waves \cite{CZG03}. 

From the theoretical point of view, one can approach the problem of current relaxation by 
evaluating the survival probability $P(t)$ of a wave packet which is initially localized 
inside an open sample of volume $\Omega$
\begin{equation}
\label{survprob}
P(t)=\int_{\Omega} |\Psi(t,{\bf r})|^2 d{\bf r} \approx \int_0^{\infty}d\Gamma \Gamma 
{\cal P}(\Gamma) \exp(-\Gamma t).
\end{equation}
The approximation above (modal approximation) valid \cite{M00,CZG03,CMS00,OWKG01,BCMS00}
for times larger that the Heisenberg time and allow us to calculate a dynamical quantity 
such as $P(t)$, based on information about resonances. The total current leaking out of the 
sample is then related to the survival probability by 
\begin{equation}
\label{current}
J(t)=-{\partial P(t) \over \partial t}.
\end{equation}

The ability of constructing micro-lasers with chaotic resonators which produce high-power 
directional emission \cite{NS97} as much as the experimental realizations of the so-called 
random lasers \cite{WAL95} where the feedback is due to multiple scattering within the medium 
(instead of being due to mirrors) is another reason why statistical properties of resonances 
became of fashion in our days. For the latter application the knowledge of resonance width 
distribution can result in the knowledge of the statistical properties of the lasing threshold.

The lasing threshold is given by the value of the smallest decay rate (i.e. smallest resonance 
width) of all eigenmodes in the amplification window \cite{MB98,PB99}. The underlying reasoning 
is that in the mode with the smallest decay rate the photons are created faster by amplification 
than they can leave (decay) the sample.
Assuming that the number of modes $K\gg 1$ that lies in the frequency window where the 
amplification is possible, have resonance widths $\Gamma$ that are statistically 
independent one gets for distribution of lasing thresholds ${\tilde {\cal P}} (\Gamma)$ 
\cite{MB98,PB99,P03}:
\begin{equation}
\label{laserr}
{\tilde {\cal P}} (\Gamma) = K {\cal P}(\Gamma<<1) \left[ 1- \int_0^{\Gamma} {\cal P}
(\Gamma'<<1) d\Gamma'\right]^{K-1}
\end{equation}
where we have assumed that all $K$ resonances are distributed according to ${\cal P}
(\Gamma<<1)$. The validity of this approximation was verified recently in the framework of
the RMT \cite{FM02}. An important outcome of \cite{OKG02,KOG03} was that one can identify 
in diffusive systems, traces of pre-localized states in the latter distribution and consequently 
in ${\tilde {\cal P}} (\Gamma)$. This send some light to recent experimental finding for 
random lasers which suggests the appearance of localized modes in diffusive samples \cite{CLXB02}.

%---------------------------------------------------------------------------------------
\subsection {\bf Ballistic Regime}
\label{subsec:balres}

At the ballistic regime, RMT modeling is applicable. Its main advantage is its universality. 
At the same time, universality means that the RMT modeling doesn't "know" anything about the 
specific properties of the system under study. Since no physical parameters (except the global 
symmetries) are plugged in the RMT machinery, it is clear that it can give the correct predictions 
only in some limiting cases, when all physical parameters and scales can be considered as irrelevant. 

\begin{figure}
\begin{center}
\epsfxsize.5\textwidth%
\includegraphics[width=10cm]{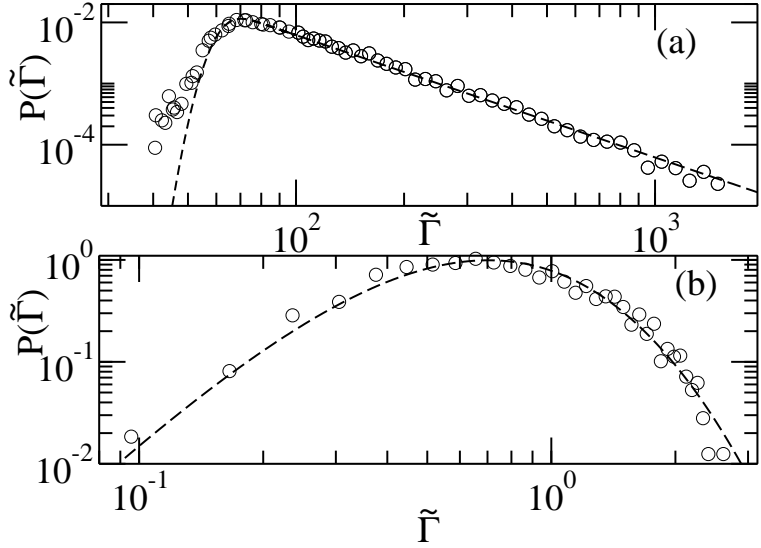}
%\epsfbox{fig1.eps}
\caption {\label{gbal} Distributions of the resonance widths in the ballistic regime for two 
different models: (a) The Open Kicked Rotor \cite{KOG03} and (b) a fully connected quantum 
graph \cite{KS00} with "generic" vertex scattering matrices. The numerical data ($\circ$) are 
in excellent agreement with the RMT predictions of Eq.~(\ref{FSpole}) (dashed lines). }
\end{center} 
\end{figure}

In the general case, Fyodorov and Sommers \cite{FS97} proved that the distribution of scaled 
resonance widths ${\tilde \Gamma}=\Gamma/\Delta$ for the unitary random matrix ensemble, is given by
\begin{equation}
\label{FSpole}
{\cal P}({\tilde \Gamma}) = \frac {(-1)^M}{\gamma(M)} {\tilde \Gamma}^{M-1} {d^M\over 
d{\tilde \Gamma}^M}
\left({\rm e}^  {-{\tilde \Gamma} \pi q} {\sinh({\tilde \Gamma}\pi)\over({\tilde \Gamma}\pi)}\right),\quad 
q={1+|\langle S\rangle|^2 \over 1-|\langle S\rangle|^2}
\end{equation}
where the parameter $q$ controls the degree of coupling with the channels, $\langle \cdots 
\rangle$ indicates an average over realizations and $\gamma(.)$ is the $\gamma-$function. In 
Fig.~\ref{gbal} we report some representative 
results from two models in the ballistic regime together with the theoretical prediction 
(\ref{FSpole}). The excellent agreement is evident.

%------------------------
%------------------------

In the limit of $M\gg 1 $, Eq.~(\ref{FSpole}) reduces to the following expression \cite{FS97}
\begin{equation}
\label{gam-rmt}
{\cal P}({\tilde \Gamma})=
\left\{
\begin{array}{lll}
{M\over 2\pi {\tilde \Gamma}^2}&\ {\rm, for}\ &{M\over \pi(q+1)}<
{\tilde \Gamma}<{M\over\pi(q-1)} \\
\ 0&  {\rm, otherwise} &
\end{array}
\right.  \ .
\end{equation}

The following argument provide some intuition about the form of resonance width distribution
(\ref{gam-rmt}). First we need to recall that the inverse of $\Gamma$ represents the quantum 
lifetime of a particle in the corresponding resonant state escaping into the leads. Moreover 
we assume that the particles are uniformly distributed inside the sample and spread ballistically
until they reach the boundary, where they are absorbed. Then we can associate the corresponding 
lifetimes with the time $t_R\sim 1/ \Gamma_R $ a particle needs to reach the boundaries, when 
starting a distance $R$ away. This classical picture can be justified for all states with 
$\Gamma \gtrsim \Gamma_{\rm cl}\gg \Delta$ 
where $\Gamma_{\rm cl}$ is the classical decay rate 
which can be calculated numerically from the exponential decay of the classical probability to stay 
inside the sample. (For RMT models we have the so-called Moldauer-Simonius relation 
$\Gamma_{\rm cl} \sim \Delta M \ln(1-|\langle S\rangle |^2)$ which give us the lower bound in 
Eq.~(\ref{gam-rmt}) while for generic chaotic system $\Gamma_{\rm cl}\sim \hbar (s/\Omega) v$ where
$s$ is the width of the opening, $\Omega$ the total volume of the system and $v$ the velocity
of the particle moving inside the system). The relative number of states that require a time 
$t<t_R$ in order to 
reach the boundaries (or equivalently the number of states with $\Gamma >\Gamma_R$) of a 
$d-$dimensional system with linear dimension $L$ is
\begin{equation}
\label{igam}
{\cal P}_{\rm int}(\Gamma_R)\equiv \int_{\Gamma_R}^{\infty}{\cal P}(\Gamma)d\Gamma
\sim\frac{V(t_R)}{L^d} = \frac {L^d-(L-R)^d}{L^d}
\end{equation}
where $V(t_R)\sim L^d-(L-R)^d$ is the volume populated by all particles with lifetimes 
$t<t_R$. Assuming ballistic motion i.e. $R=v t_R$, 
we get from Eq.~(\ref{igam}) in the limit where $\Gamma\gg \Gamma_{\rm cl}$
\begin{equation}
\label{obound}
{\cal P}_{\rm int}(\Gamma_R)\sim {1\over \Gamma_R}
\end{equation}
which eventually leads to the RMT prediction Eq.~(\ref{gam-rmt}).

Eqs.~(\ref{FSpole},\ref{gam-rmt}) is our starting point. In the next subsections we will 
investigate how deviations from these expressions raise as we increase the randomness of the
system.

%---------------------------------------------------------------------------------------
\subsection {\bf Diffusive Regime}
\label{subsec:difres}

\begin{figure}
\begin{center}
\epsfxsize.7\textwidth
\includegraphics[width=8cm]{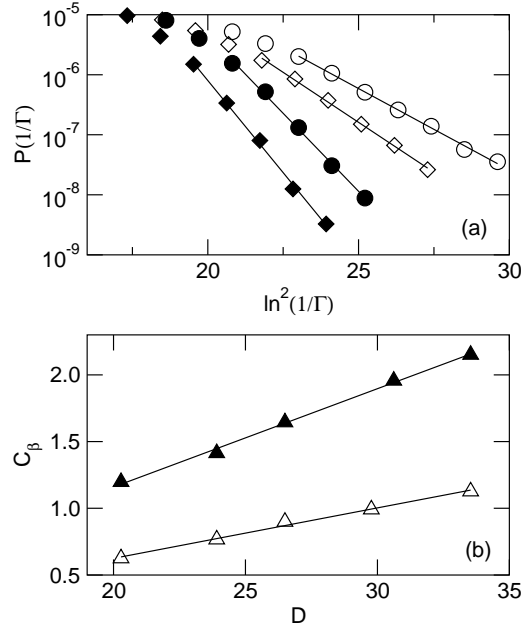}
%\epsfbox{fig2.eps}
\caption {\label{dif1} (a) The distribution of resonance widths (plotted as ${\cal P}(1/\Gamma)$
vs. $1/\Gamma$) for $\Gamma < \Gamma_{cl}$ for two representative values of $D$ for the 2-
dimensional Kicked Rotator in the diffusive regime \cite{KOG03,OKG03}. The system size in all 
cases is $L=80$. Filled symbols correspond to broken TRS. The solid lines are the best fit of 
Eq.~(\ref{difgammaS}) for $\beta = 1 (2)$ to the numerical data. (b) Coefficients $C_{\beta}$ 
vs. $D$ for the same model as in (a) \cite{KOG03,OKG03}. The solid lines are the best fits to 
$C_{\beta}= A_{\beta} D+ B_{\beta}$ for $\beta=1 (2)$. The ratio $R=A_2/A_1 = 1.95\pm 0.03$. } 
\end{center}
\end{figure}

We start our presentation with the study of small-resonance width distribution ${\cal P}
(\Gamma < \Delta)$. The small resonances $\Gamma < \Delta$ can be associated, with the 
existence of pre-localized states of the closed system (for a discussion on pre-localized 
states see \cite{M00,OKG02,KOG03,UMRS00}). They consist of a short-scale bump (where most of the norm 
is concentrated) and they decay rapidly in a power law fashion from the center of localization 
\cite{M00,FE95,KOG03,UMRS00}. One then expects that states of this type with localization 
centers at the bulk of the sample are affected very weakly by the opening of the system 
at the boundaries. In first order perturbation theory, considering the opening as a small 
perturbation we obtain \cite{KOG03,OKG03}
\begin{equation}
\label{pertgamma}
{\Gamma\over 2} = \langle \Psi|V^{\dagger}V|\Psi\rangle =\sum_{n\in {\rm boundary}}|\Psi(n)|^2 \sim 
L^{d-1} |\Psi(L)|^2 
\end{equation}
where $|\Psi (L)|^2$ is the wavefunction intensity of a pre-localized state at the boundary 
and $d$ is the dimensionality of the sample. At the same time the distribution of wavefunction 
components at the boundary was found to be \cite{FE95}
\begin{equation}
\label{theta}
{\cal P}(\theta) \sim \exp\left(-A^{(d)}_{\beta}\ln^d (\theta^{4-d})\right),\quad 
\theta=1/L^{(d-1)/2} \Psi(L)
\end{equation}
where the coefficient $A^{(d)}\propto \beta D$. Here $\beta=1(2)$ denotes the corresponding 
ensemble for preserved (broken) time-reversal symmetry. Using Eq.~(\ref{theta}) together with 
Eq.~(\ref{pertgamma}) we obtain 
\begin{equation}
\label{difgammaS}
{\cal P}(\Gamma<\Delta) \sim \exp\left(-C^{(d)}_{\beta} \ln^d (1/\Gamma)\right),\quad {\rm where} 
\quad C_{\beta} \propto \beta D
\end{equation}
A detailed numerical analysis performed in \cite{KOG03,OKG03,WMK05} for the $d=2,3$ showed 
that the above perturbative derivation valids as well for relatively large resonances i.e.  
$\Delta < \Gamma < \Gamma_{cl} $ where $\Gamma_{cl} = D/L^2$ is the inverse Thouless time.
In Fig.~\ref{dif1} we report some numerical data for the case $d=2$ and compare with the
theoretical predictions of Eq.~(\ref{difgammaS}). Let us finally compare Eq.~(\ref{difgammaS}) 
with the results for ballistic systems (see Eq.~(\ref{gam-rmt})). In the latter case, a strip 
free of resonances is formed. In the diffusive regime, in contrast, there are pre-localized 
states, which are weakly coupled to the leads. Due to their existence the distribution of 
the small resonance widths has a non-trivial behavior described by Eq.~(\ref{difgammaS}).  

%--------------------
%--------------------

Next we turn to the analysis of ${\cal P}(\Gamma)$ for $\Gamma \gtrsim \Gamma_{cl}$. Using 
the same argument that led to Eq.~(\ref{igam}), but assuming now diffusive spreading i.e.
$R^2=D\times t$, we get (to leading order approximation with respect to $\Gamma_{\rm cl}/\Gamma$)
\begin{equation}
\label{gamlargedif}
{\cal P}(\Gamma)\sim \left(\frac{1}{\Gamma}\right)^{\frac{3}{2}}.
\end{equation}
valid for quasi-one \cite{BGS91}, two \cite{OKG02,OKG03} or three-dimensional \cite{WMK05} random
media as long as the leads are attached to the boundary of the sample. We conclude that the different 
power law decay of Eq.~(\ref{gamlargedif}) with 
respect to Eqs.~(\ref{gam-rmt},\ref{obound}) is a result of the different nature of the 
dynamics: ballistic vs. diffusive.

Here it is interesting to point that a different way of opening the system might lead 
to a different power law behavior for ${\cal P}(\Gamma)$. Such a situation can be 
realized if instead of opening the system at the boundaries we introduce "one-site" 
absorber (or one "lead") somewhere inside the sample. In such a case for $d=2$ we have 
\begin{equation}
\label{finalg1}
{\cal P}_{\rm int}(\Gamma_R)\sim\frac{V(t_R)}{L^2}=\frac{R^2}{L^2}=\frac{Dt_R}{L^2}\sim
\frac{\Gamma_{cl}}{\Gamma_R}
\end{equation}
leading to the following power law decay
\begin{equation}
\label{finalg1a}
{\cal P}(\Gamma)\sim (\frac{1}{\Gamma})^2.
\end{equation}
Similarly, the analog of Eq.~(\ref{finalg1a}) in $d=3$ is 
\begin{equation}
\label{finalg2}
{\cal P}(\Gamma)\sim (\frac{1}{\Gamma})^{2.5}
\end{equation}
where we had to substitute $V(t_R)\sim R^3$. The above results valid for any number $M$ 
of "leads" such that the ratio $M/L^d$ scales as $1/L^d$. 

If on the other hand we attach the open channel to the boundary (assume square geometry) of 
a $3d$ sample we come out with a decay law which is the same as the one given by 
Eq.~(\ref{finalg1a}). This is due to the fact that the decay from the surface leads to a 
situation alike the one of a $2d$ system.

\begin{figure}
\begin{center}
\epsfxsize.5\textwidth%
\includegraphics[width=10cm]{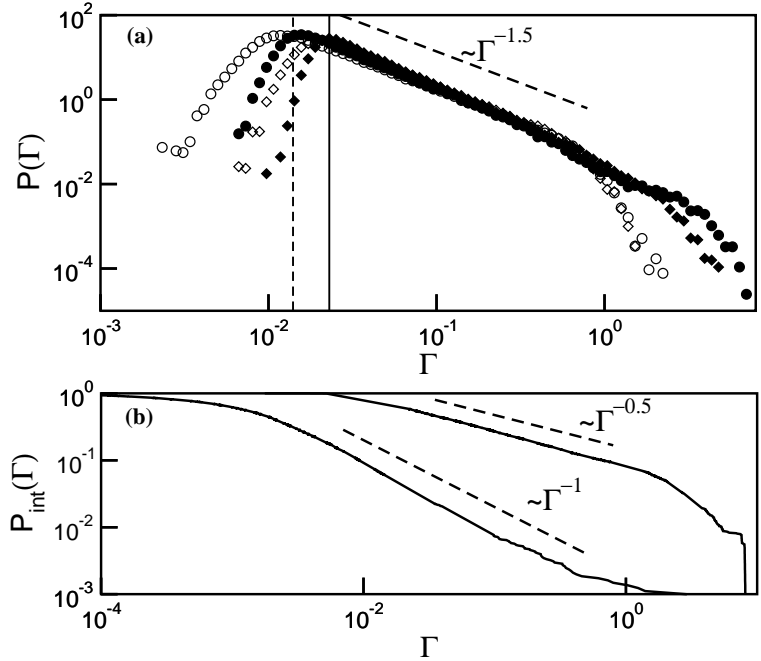}
%\epsfbox{fig3.eps}
\caption {\label{dif2}(a) The resonance width distribution ${\cal P}(\Gamma)$ for preserved
TRS and $D=20.3$ ($\circ$) and $D=33.5$ ($\diamond$). The corresponding filled symbols
represent ${\cal P}(\Gamma)$ for broken TRS and the same values of $D$. The dashed 
(solid) vertical line mark the classical decay rate $\Gamma_{cl}$  for 
$D=20.3 (D= 33.5)$. (b) The ${\cal P}_{\rm int} (\Gamma)$ for a sample with nine leads 
(lower curve). For comparison we plot also the ${\cal P}_{\rm int}(\Gamma)$ for the same 
sample but when we open the system from the boundaries. The dashed lines correspond 
to the theoretical predictions (\ref{gamlargedif}) and (\ref{finalg1a}). The figure is 
taken from \cite{OKG03}.}
\end{center}
\end{figure}

\begin{figure}
\begin{center}
\epsfxsize.5\textwidth%
\includegraphics[width=10cm]{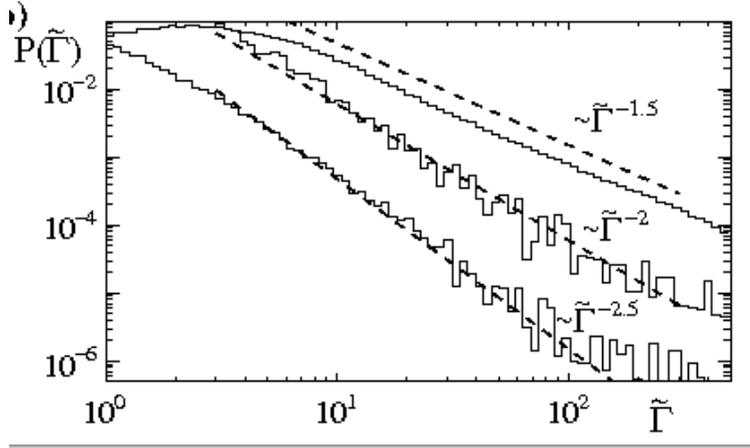}
%\epsfbox{fig4.ps}
\caption {\label{dif3}(a) The resonance width distribution ${\cal P}({\tilde \Gamma})$ for
the $3d$ Anderson model \cite{WMK05} and various configurations of the open channels.
The dashed lines are the corresponding theoretical predictions given by 
Eqs.~(\ref{gamlargedif},\ref{finalg1a},\ref{finalg2}), (see text). The figures is taken from 
\cite{WMK05}.
}
\end{center}
\end{figure}

In Fig.~\ref{dif2} we present some numerical calculations of the $2d$ Kicked Rotator \cite{OKG03} 
while in Fig.~\ref{dif3} we present numerical data from the $3d$ Anderson model in the diffusive 
regime \cite{WMK05}. In all cases a comparison with the corresponding theoretical predictions 
(\ref{gamlargedif},\ref{finalg1a},\ref{finalg2}) shows a nice agreement. 
%--------------------
%--------------------

%--------------------
%--------------------

%-------------------------------------------------------------------------------------------------
\subsection {\bf Localized Regime}
\label{subsec:locres}

Various groups \cite{WMK05,TF00,SJNS00,PROT04} had investigated the resonance width distribution 
in the localized regime during the lasts years. 

In the region of exponentially narrow resonances $\Gamma<\Gamma_0=\exp(-2L/\xi)$ the 
distribution was found to be log-normal i.e.
\begin{equation}
\label{locres2}
{\cal P}(\tilde \Gamma)\sim \exp[-(4 {L\over \xi})^{-1} \ln^2(\tilde \Gamma)] ,
\quad  \Gamma< \Gamma_0. 
\end{equation}
This is entirely analogous to the conductance distribution 
of localized systems. Equation (\ref{locres2}) essentially relies on two assumptions: first, 
that eigenfunction components are randomly distributed with no long-range correlations, and 
second, that they are exponentially localized with a normal distribution of localization lengths.
This part of the distribution becomes negligible at large $L$, because it comes of a fraction 
$\sim \xi/L$ of the full set of all resonances.

Instead the long resonance tails behave as 
\begin{equation}
\label{locres1}
{\cal P}({\tilde \Gamma})\sim \left({\xi\over L}\right)
{1\over {\tilde \Gamma}},\quad  \Gamma_0< \Gamma \ll 1/L
\end{equation}
Eq.~(\ref{locres1}) can be easily understood once employing Eq.~(\ref{igam}). The new ingredient 
now is that wavefunctions are exponentially localized i.e. $|\Psi(r)|\sim (1/\xi^{d/2})
\exp(-r/\xi)$. Using simple perturbation arguments, we have that (see Eq.~(\ref{pertgamma}))
$\Gamma \sim |\Psi(r)|^2$ which leads to the following approximation about the volume $V(t_R)
\propto R^d \sim \xi^d \ln^d(\xi^d\Gamma)$. Inserting
this in Eq.~(\ref{igam}) we get (to leading order approximation with respect to $\xi/L$)
Eq.~(\ref{locres1}). 

The large $\Gamma$ region is essentially determined by the coupling to continuum, so it should 
be model-dependent. Nevertheless, it is reasonable to assume that the number of resonances involved 
is constant, of order $\xi$ and therefore this extreme tail should subside at large $L$,
at a rate $\sim \xi/L$.

\begin{figure}
\begin{center}
\epsfxsize.5\textwidth%
\includegraphics[width=10cm]{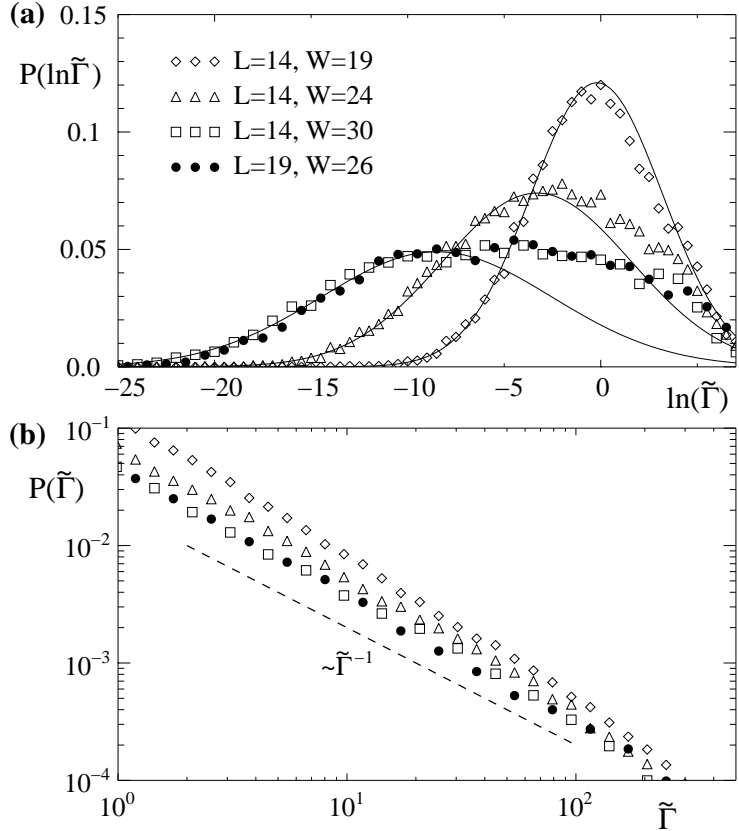}
%\epsfbox{fig5.eps}
\caption {\label{loc}${\cal P}(\tilde{\Gamma})$ in the localized regime for various combinations
of $V$ and $L$ in the range $\tilde{\Gamma}\le 1$. The log-normal decay is highlighted
by Gaussian fits (full curves) whose maximum decreases increasing
strength of disorder and also shifts towards smaller values of $\tilde{\Gamma}$.
Keeping the ratio $\xi/L\approx0.136$ fixed, coinciding
distributions (filled circles and open squares) are obtained for
different combinations of $L,\,V$. (b) For $\tilde{\Gamma}\ge 1$ the anticipated
power-law decay ${\cal P}(\tilde{\Gamma})\sim 1/\tilde{\Gamma}$ is
observed (dashed line) which  becomes more robust for increasing strength of disorder.
The figure is taken from \cite{WMK05}.
}
\end{center}
\end{figure}

From Eqs.~(\ref{locres2},\ref{locres1}) it becomes evident that ${\cal P}(\tilde{\Gamma})$ depends
on one parameter. Namely the dimensionless parameter $\xi/L$. This dimensionless parameter
is the cornerstone of the one-parameter scaling theory of localization. It was shown in the past
that the dimensionless conductance $g$ is a simple function of $\xi/L$. The above theoretical 
considerations were tested \cite{WMK05,SJNS00,PROT04} for various disordered and chaotic systems 
with dynamical localizations and were found to describe nicely the numerical data. In Fig.~\ref{loc} 
we report some representative cases from the 3D Anderson model \cite{WMK05} in the localized regime. 

Let us finally note that in the thermodynamic limit $L\rightarrow \infty$ the probability of
finding an eigenstate at any finite distance from the boundary is equal to zero. Thus the
distribution of the resonance widths in this case collapses into a delta function centered
at zero.

%--------------------
%--------------------

%-------------------------------------------------------------------------------------------------
\subsection {\bf Criticality}
\label{subsec:critres}

\begin{figure}
\begin{center}
\epsfxsize.5\textwidth%
\includegraphics[width=10cm]{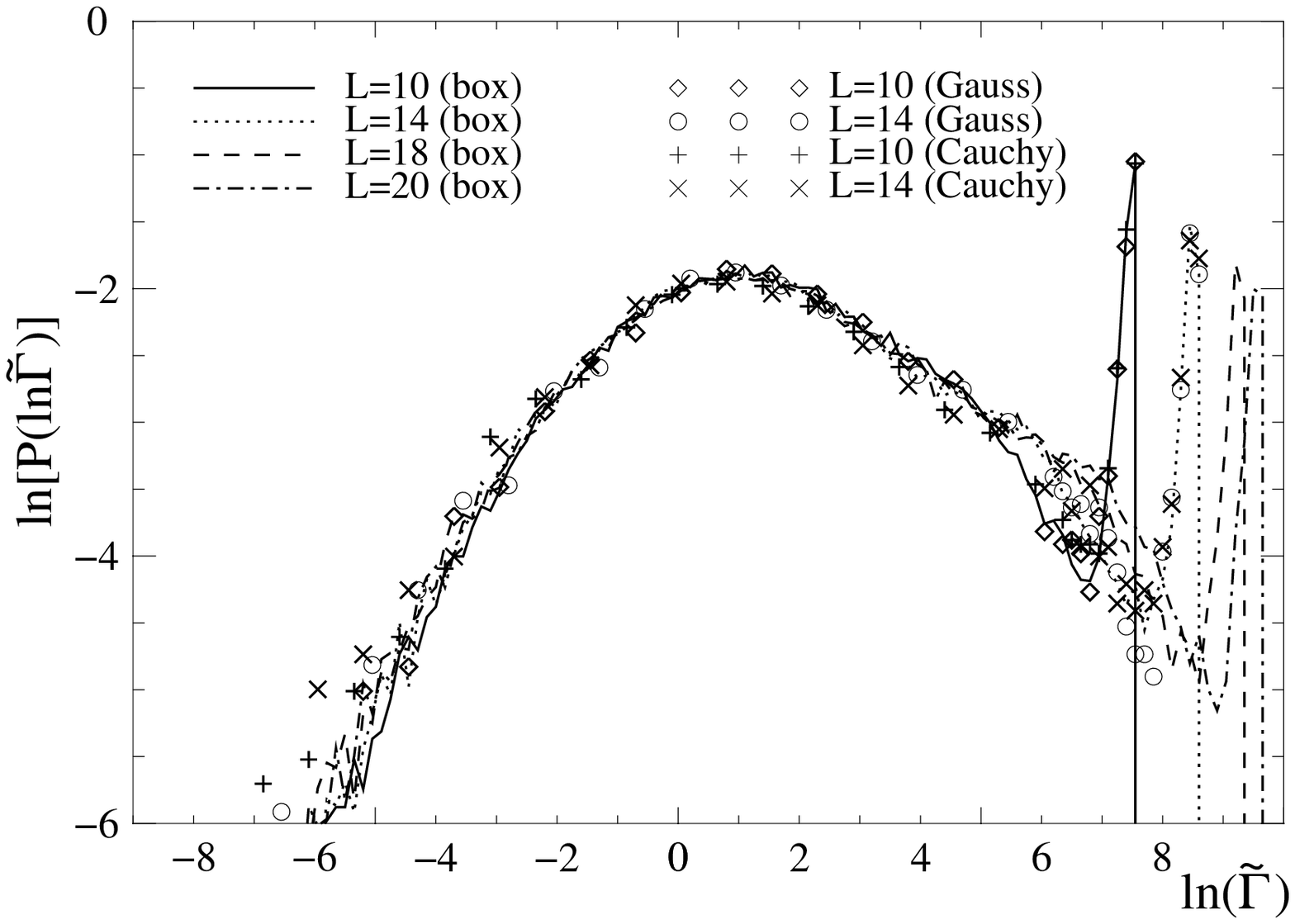}
%\epsfbox{fig6_0.eps}
\caption{\label{crit1_0}
Universal behavior of ${\cal P}({\tilde \Gamma})$ at the MIT for a $3d$ Anderson model. The
figure is taken from \cite{KW02}}.
\end{center}
\end{figure}

The statistical properties of various observables at the MIT is one of the most intriguing
problems for many years now. Actually, despite the rich activity \cite{A58,M00,FE95,AS86,W80,EM00,BHMM01}
very few theoretical results are known. Here, we present consequences of the MIT on the 
statistical properties of the rescaled resonance widths ${\tilde \Gamma}$. It was found 
\cite{KW02} that ${\cal P}({\tilde \Gamma})$ follow a new universal distribution, i.e. 
independent of the microscopic details of the random potential, and number of channels $M$ 
as can be seen from Fig.~\ref{crit1_0}. For small resonance widths i.e.  ${\tilde \Gamma}<1$, 
we have found \cite{WMK05} that ${\cal P}({\tilde \Gamma})$ can be fitted with a log-normal. 
The sharp peak on the extreme right corresponding to very large resonances is non-universal 
(model specific) and statistically insignificant since it subside as $L$ increases like $M/L^3 
\sim L^{-1}$ (see also discussion at the previous section).

%--------------------------------
%--------------------------------

On the other hand, the main part of ${\cal P}({\tilde \Gamma})$, corresponding to intermediate 
large resonances, follow a power-law which is different from those found for ballistic, diffusive 
or localized systems (see previous sections) i.e.
\begin{equation}
\label{MITgam}
{\cal P}({\tilde \Gamma})\sim g_{\rm c}^{1/d}  {\tilde \Gamma}^{-(1+1/d)}.
\end{equation}
where $g_{\rm c}$ is the Thouless conductance at criticality.

%--------------------------------
\begin{figure}
\begin{center}
\epsfxsize.5\textwidth%
\includegraphics[width=10cm]{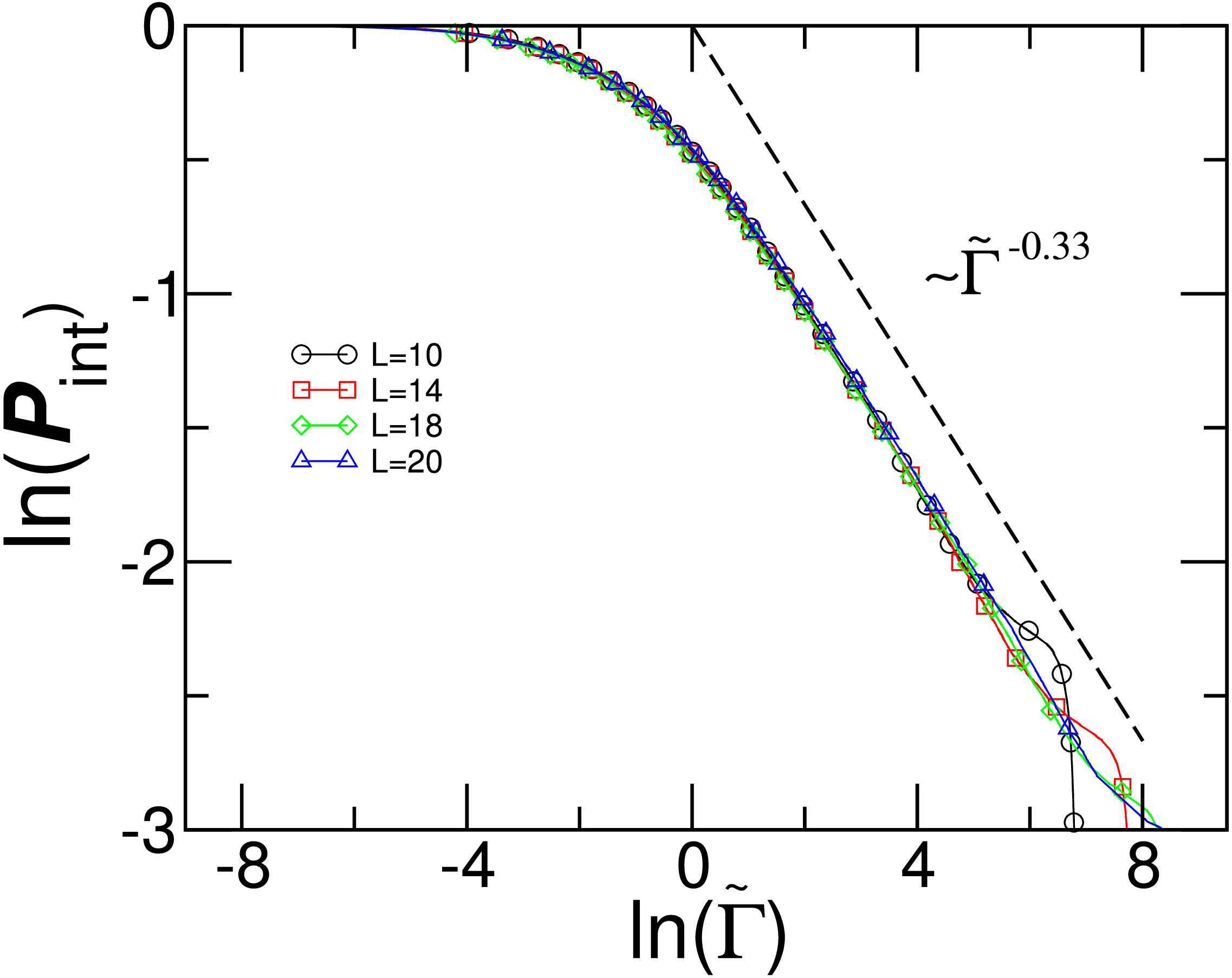}
%\epsfbox{fig6.ps}
\caption{\label{crit1}
The integrated distribution ${\cal P}_{\rm int}({\tilde \Gamma})$ for the $3d$
Anderson model at MIT \cite{KW02,WMK05}. The dashed line is the theoretical prediction
(\ref{MITgam}) corresponding to ${\cal P}_{\rm int}({\tilde \Gamma}) \sim {\tilde 
\Gamma}^{-0.333}$ for our case. The figure is taken from \cite{WMK05,KW02}.}
\end{center}
\end{figure}
%--------------------------------

One can relate the power-law decay (\ref{MITgam}) to the anomalous diffusion at the MIT. 
Indeed, at MIT the conductance of a $d$-dimensional disordered sample has a finite value 
$g_{\rm c}\sim 1$. Approaching the MIT from the metallic side one has $g\sim E_T/ 
\Delta$, where $E_T=D/R^2$ is the Thouless energy, $D$ is the diffusion coefficient, 
and $\Delta \sim 1/R^d$ is the mean level spacing in a $d$-dimensional sample with 
linear size $R$. This yields $D\sim g_{\rm c}/R^{d-2}$ at $W_c$. Taking into
account that $D=R^2/t$, we get for the spreading of an excitation at the MIT
\begin{equation}
\label{trpert}
R^d(t)\sim g_{\rm c} t\,\,.
\end{equation}
A straightforward application of Eq.~(\ref{igam}) then, leads to Eq.~(\ref{MITgam}). 
In Fig.~\ref{crit1} we report some numerical results for the 3D Anderson model at MIT
\cite{WMK05,KW02}. An inverse power law ${\cal P}_{ \rm int}({\tilde \Gamma}) \sim {\tilde 
\Gamma}^{ -\alpha}$ is evident. The best fit to the numerical data yields $\alpha=
0.333\pm 0.005$ in accordance with Eq.~(\ref{MITgam}). 

%--------------------------------
\begin{figure}
\begin{center}
\epsfxsize.5\textwidth%
\includegraphics[width=10cm]{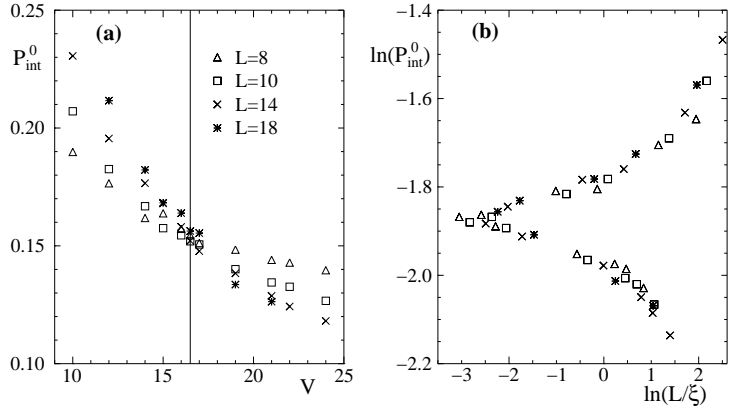}
%\epsfbox{fig7.eps}
\caption{\label{crit2} (a) ${\cal P}_{\rm int}^0(W,L)$ as a function of disorder strength 
$W$ for different 
system sizes $L$ provides a means to determine the critical point $W_c$ of the MIT 
(vertical line at $W_c=16.5$). (b) The one-parameter scaling of ${\cal P}_{\rm int}^0(W,L)$ 
[Eq.~(\ref{scale})] is confirmed for various system sizes $L$ and disorder strengths $W$
using the box distribution. 
The figure is taken from \cite{WMK05,KW02}.}
\end{center}
\end{figure}
%--------------------------------
%

In the original proposal of the scaling theory of localization, the conductance $g$ 
is the relevant parameter \cite{A58}. A manifestation of this statement is seen in 
Eq.~(\ref{MITgam}) where ${\cal P}_{\rm int}^{0}\equiv {\cal P}_{\rm int}({\tilde \Gamma}_0
)$ is proportional to the conductance $g$. It is therefore natural to expect that 
${\cal P}_{\rm int}^{0}$ will follow a scaling behavior for finite $L$ (and for some 
${\tilde \Gamma}_0\sim 1$), that is similar to the one obeyed by the conductance $g$. 
It was therefore postulated in \cite{WMK05,KW02} the following scaling hypothesis
\begin{equation}
\label{scale}
{\cal P}_{\rm int}^{0}(W,L)=f(L/\xi(W))\,\, ,
\end{equation}
where $\xi(W)$ is the correlation length at MIT. In the insulating phase 
($W>W_c$) the conductance of a sample with length $L$ behaves as $g(L)\sim 
\exp(-L/\xi)$ due to the exponential localization of the eigenstates, and 
therefore we have $g(L_1) < g(L_2)$ for $L_1>L_2$. Based on Eq.~(\ref{MITgam}) we expect 
the same behavior for ${\cal P}_{\rm int}^{0}$ i.e. for every finite $L_1>L_2$ we must 
have ${\cal P}_{\rm int}^{0}(W,L_1)<{\cal P}_{\rm int}^{0}(W,L_2)$. On the other 
hand, in the metallic regime ($W<W_c$) we have that $g(L)=D L^{d-2}$ and therefore 
for $d>2$, we expect from Eq.~(\ref{MITgam}) ${\cal P}_{\rm int}^{0}(W,L_1)>{\cal P}_{\rm int}^{0}
(W,L_2)$. Thus, the critical 
point is the one at which the size effect changes its sign, or in other words, 
the point where all curves ${\cal P}_{\rm int}^{0}(W,L)$ for various $L$ cross.
One can reformulate the last statement by saying that in the thermodynamic limit 
$L\rightarrow \infty$ at $W=W_c$ the number of resonances with width larger 
than the mean level spacing goes to a constant. 

In Fig.~\ref{crit2}a, we show the evolution of ${\cal P}_{\rm int}^{0}(W)$ for 
different $L$ using the box distribution \cite{WMK05,KW02}. From this analysis
the critical disordered strength $W=W_c=16.5\pm 0.5$ was determined in \cite{WMK05,KW02}
in agreement with other calculations \cite{A58,RMS01}. A further verification
of the scaling hypothesis (\ref{scale}) is shown in Fig.~\ref{crit2}b where the 
same data are reported as a function of the scaling ratio $L/\xi$. All points collapse 
on two separate branches for $W<W_c$ and $W>W_c$. 

%=======================================================================================
\section {\bf Delay times}
\label{sec:delay}

We turn now to the analysis of Wigner and proper delay times as defined in Eq.~(\ref{Wtau}) 
above. Their knowledge is relevant for experiments on frequency and parameter-dependent 
transmission through chaotic microwave cavities \cite{DS90,KKS95} or semiconductor quantum 
dots with ballistic point contacts \cite{W97}. Also, it can be shown that they are related 
to the distribution of reflection coefficients $R$ in the present of weak absorption. 

Absorption is one of the main ingredients in actual experimental situations and gain a lot 
of interest the last years (see \cite{SFS05} and \cite{KSW05} in this volume). Unfortunately 
a comprehensive treatment of absorption is still lacking. There are only very few reported 
analytical results for the distribution ${\cal P}(R)$ of the reflection coefficient $R=
S^{\dagger}S$ in the 
presence of absorption and all of them are within the regime of applicability of RMT 
\cite{SFS05,KML00,SS03}, except the recent work \cite{F03,MB96}, where quasi-1D geometry in the 
localized regime is considered as well. 

Specifically, in the weak absorption limit it was shown \cite{DS90,KML00} that the following 
relation holds:
\begin{equation}
\label{rcoef}
R_q = 1-\tau_q/\tau_a
\end{equation}
where $\tau_q$ are the proper delay times (eigenvalues of the Wigner-Smith operator) and
$1/\tau_a$ is the absorption rate. Thus the knowledge of ${\cal P}(R)$ reduces to the
calculation of the distribution of proper delay times ${\cal P}(\tau_q)$ \cite{KML00}.

For one channel the distribution of delay times is now quite well understood and studied in
all regimes. Quite recently it was shown \cite{OF05} the existence of a very general
relation between the delay time distribution and the distribution of eigenfunction intensities 
\begin{equation}
\label{9a}
\left\langle \tilde{\tau}_w^{-k}\right\rangle=\left\langle
y^{k+1}\right\rangle
\end{equation}
where $y=\Omega|\psi_n(r)|^2$ is the local eigenfunction intensity and $\tilde{\tau}=\tau\Delta/2\pi$. 
Eq.~(\ref{9a}) leads to the following functional relation between the two distributions:
\begin{equation}
\label{tau_wave}
\tilde{\cal P}_{w}
(\tilde{\tau}_w)=\frac{1}{\tilde{\tau}_w^3}{\cal P}_y
\left(\frac{1}{\tilde{\tau}_w}\right).
\end{equation}
On one hand, this relation allows us to use the existent knowledge on
eigenfunction statistics \cite{M00} to provide explicit expressions for delay
times distributions in various regimes of interest. On the other
hand, since phase shifts and delay times are experimentally
measurable quantities, especially in the one-channel reflection
experiment \cite{DS90,KML00,KMMS04,JPM03,GSST82}, this relation opens
a new possibility for experimental study of eigenfunctions.

%----------------------------------------------------------------------------------
\subsection {\bf Ballistic Regime}
\label{sec:delayball}

We start again our presentation from the ballistic regime. The notion of proper delay 
times goes back more than 40 years to the seminar paper of Smith \cite{W55}. Although
many authors have worked on this problem \cite{FS97,GMB96,LW91,SZZ96,BB97,ISS94}, only recently 
its probability distribution was calculated. It was shown in \cite{BFB97} using standard RMT 
methods, that the distribution of inverse proper delay times is given by the Laguerre ensemble 
from random matrix theory
\begin{equation}
\label{tauball}
{\cal P}(\tau_1^{-1},\cdots,\tau_M^{-1})\propto \prod_{i<j}|\tau_i^{-1}-\tau_j^{-1}|^{\beta}
\prod_{k}|(\tau_i^{-1})^{\beta M/2} \exp^{-\beta 2\pi \hbar\tau_k^{-1}/2\Delta}.
\end{equation}

As a matter of fact from Eq.~(\ref{tauball}) one can evaluate the distribution
of Wigner delay times $\tau_W$, which for large values decays as a power-law
\begin{equation}
\label{tauballlong}
{\cal P}(\tau_W)\propto {1\over \tau_W^{2+\beta M/2}}
\end{equation}
in agreement with an earlier conjecture by Fyodorov and Sommers \cite{FS97}.

Specifically for $M=1$ the distribution of Wigner delay times was calculated even in the
crossover regime between unitary ($\beta=2$) and orthogonal ($\beta=1$) symmetry classes
and was found to be \cite{FSS97}
\begin{eqnarray}
\label{cross}
{\mathcal P}_{w}(\tilde{\tau}_w)&=&\frac{1}{2\td^3}\int_{-1}^{1}d\lambda\;
\int_{1}^{\infty}d\lambda_2\:\lambda_2^2 e^{-X^2(\lambda_2^2-1)}e^{-
\lambda_2^2/\td}I_0\left[\frac{\lambda_2\sqrt{\lambda_2^2-1}}{\td}\right]
{\cal T}_2(\lambda,\lambda_2),\\
{\cal T}_2(\lambda,\lambda_2)&=&
2X^2\left[(1-\lambda^2)e^{-\alpha}+\lambda_2^2
(1-e^{-\alpha})\right]-(1-e^{-\alpha}),\nonumber\\
\end{eqnarray}
where $\alpha=X^2(1-\lambda^2)$, $I_0(z)$ stands for the modified Bessel function, and 
$X$ is a crossover driving parameter. For pure symmetries Eq.~(\ref{cross}) leads to
\cite{FS97,OF05,GMB96} 
\begin{equation}
\label{tauballlong1}
{\cal P}(\tau_W)= [(\beta/2)^{\beta/2}/\gamma(\beta/2)]\tau_W^{-\beta/2 -2}
\exp^{-\beta/2\tau_W}.
\end{equation}

The following simple argument can be used in order to understand the tails of 
the  Wigner delay-time distribution Eq.~(\ref{tauballlong}). Our starting point 
is the well known relation
\begin{equation}
\label{dtime}
\tau(E)=\sum_{n=1}^{M} \frac{\Gamma_n}{(E-E_n)^2 + \Gamma_n^2/4} 
\end{equation}
which connects the Wigner delay times and the poles of the $S-$matrix. 

%------------------------
\begin{figure}
\begin{center}
\epsfxsize.5\textwidth%
\includegraphics[width=10cm]{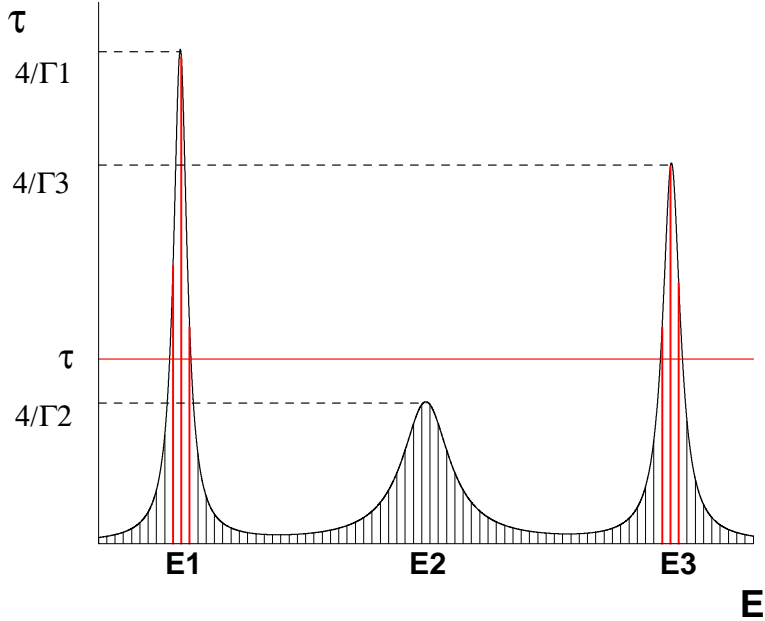}
%\epsfbox{fig8.eps}
\caption {\label{wign1} 
Schematic plot for the Wigner delay time as a function of energy according to
Eq.~(\ref{dtime}).
}
\end{center}
\end{figure} 
%------------------------

It is evident that large times $\tau(E)\sim \Gamma^{-1}_n$ corresponds to the cases when 
$E\simeq E_n$ and $\Gamma_n \ll 1$. In the neighborhood of these points, $\tau(E)$ can be 
approximated by a single Lorentzian (\ref{dtime}). Sampling the energies $E$ with step 
$\Delta E\ll \Gamma_{min}$ we calculate the number of points for which the time delay is 
larger than some fixed value $\tau$ (see Fig.~\ref{wign1}). Assuming that the contribution 
of each Lorentzian is proportional to its width one can estimate this number as $\sum_{
\Gamma_n < 1/\tau}\Gamma_n/\Delta E$. For the integrated distribution of delay times in 
the limit $\Delta E \rightarrow 0$ we obtain  
\begin{equation}
\label{argtau}
{\cal P}_{int} (\tau) \sim \int^{1/\tau} d\Gamma {\cal P}
(\Gamma)\Gamma 
\end{equation}
and by substituting the small resonance width asymptotic given by Eq.~(\ref{FSpole})
we come out with the power-law expression ~(\ref{tauballlong}). 

%----------------------------------------------------------------------------------
\subsection{\bf Diffusive regime}

Using the general relation (\ref{tau_wave}) for the case of $M=1$, 
we find for ${\cal P}(\tau_W)$ \cite{OF05} 
\begin{eqnarray}
\label{metallic}
{\mathcal P}_{w}(\tilde{\tau}_w)&=&\frac{e^{-1/2\td}}{\sqrt{2\pi}\td^{5/2}}
\left[1+\frac{\kappa}{2}\left(\frac{3}{2}-\frac{3}{\td}+\frac{1}{2\td^2}\right)+
\dots\right]\;\;\; \beta=1,\nonumber\\
\Pt &=&\frac{e^{-1/\td}}{\td^3}
\left[1+\frac{\kappa}{2}\left({2}-\frac{4}{\td}+\frac{1}{\td^2}\right)+\dots
\right]\;\;\; \beta=2,\nonumber\\
\Pt &=&\frac{4e^{-2/\td}}{\td^4}
\left[1+\frac{\kappa}{2}\left({3}-\frac{6}{\td}+\frac{2}{\td^2}\right)+\dots
\right]\;\;\; \beta=4,
\end{eqnarray}
where the parameter $\kappa\propto g^{-1}$ is inversely proportional to the dimensionless 
conductance $g$. The proportionality coefficient depends essentially on the sample geometry 
and on the coordinates of the lead. Note that in the limit of $g\rightarrow \infty$ we recover
the RMT results discussed in the previous section.

Eq.(\ref{metallic}) holds for relatively large delay times $\td \gtrsim \sqrt{\kappa}$, while 
in the opposite case the distribution is dominated by the existence of the anomalously localized 
states and has the following  behavior for dimensionality $d=2,3$ 
\cite{OF05}:
\begin{eqnarray}
\Pt &\sim& \exp\left(\frac{\beta}{2}\left\{-\frac{1}{\td}+\kappa
\frac{1}{\td^2}+\dots\right\}\right), \;\;\;\; \kappa\lesssim \td
\lesssim \sqrt{\kappa},\\
\Pt &\sim& \exp({-C_d \ln^d(1/\td)}), \;\;\;\; \td \lesssim \kappa.
\end{eqnarray}
Note that although in \cite{OF05} the coefficient $C_d$ was claimed to depend only on the
dimensionality of the system, it is possible to depend also on the symmetry parameter $\beta$
as well (for a discussion on these issues see \cite{FE95,KOG03} and references therein).

\begin{figure}
\begin{center}
\epsfxsize.5\textwidth%
\includegraphics[width=10cm]{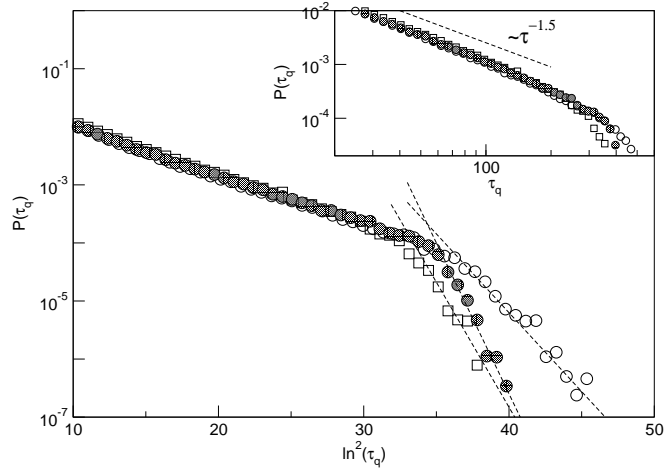}
%\epsfbox{fig9.eps}
\caption {\label{wign2}
The proper delay times distribution ${\cal P}(\tau_q)$ for a $2d$ Kicked Rotator
\cite{KOG03} with diffusive coefficients $D=20.3$ ($\circ$) and 
$D=29.8$ ($\Box$ ). The ($\bullet$) correspond to $D=20.3$ but now with broken TRS. The 
dashed lines have slopes equal to $C_{\beta}$ extracted from the corresponding ${\cal P}
(\Gamma)$ (see Fig.~8). In the inset we report ${\cal P} (\tau_q )$ for moderate values 
of $\tau_q$ in a double logarithmic scale.
The figure is taken from \cite{KOG03,OKG03}.}
\end{center}
\end{figure}

For many open channels $M\gg 1$ there are no quantitative theoretical results yet. However 
one can employ qualitative arguments which together with numerical findings can allow us
to understand the resulting distributions. Indeed, 
substituting in Eq.~(\ref{argtau}) the small resonance width asymptotic for the ${\cal P}
(\Gamma)$ given by Eq.~(\ref{difgammaS}) we come out with the following log-normal law
for the large $\tau$ regime
\begin{equation}
\label{difftausmall}
{\cal P}(\tau > \Gamma_{cl}^{-1}) \sim \exp(-C_\beta \ln^d \tau)
\end{equation}
where the coefficient $C_\beta$ is the same as the one given in Eq.~(\ref{difgammaS}).
This prediction has been tested numerically in \cite{OKG02,KOG03} for the case of a
$2d$ diffusive system, and the numerical findings (see Fig.~\ref{wign2}) were shown to be in 
excellent agreement.

Now we estimate the behavior of ${\cal P}(\tau)$ for $\tau \lesssim \Gamma_{cl}^{-1}$. 
In this regime many short-living resonances contribute to the sum (\ref{dtime}). We may
therefore consider $\tau$ as a sum of many independent positive random variables 
each of the type $\tau_n=\Gamma_n x_n$, where $x_n=\delta E_n^{-2}$. Assuming further
that $\delta E_n$ are uniformly distributed random numbers we find that the distribution
${\cal P}(x_n)$ has the asymptotic power law behavior $1/x_n^{3/2}$. As a next step we 
find that the distribution ${\cal P}(\tau_n)$ decays asymptotically as $1/\tau_n^{3/2}$ 
where we use that ${\cal P}(\Gamma_n)\sim 1/\Gamma_n^{3/2}$. Then the corresponding ${\cal P}
(\tau)$ is known to be a stable asymmetric Levy distribution $L_{\mu,1}(\tau)$ of 
index $\mu=1/2$ \cite{BG90} which has the following form  at the origin
\begin{equation}
\label{taudiforig}
{\cal P}(\tau\lesssim \Gamma_{cl}^{-1}) \sim {1\over\tau^{3/2}}\exp(-\sigma/\tau),
\end{equation}
where $\sigma$ is some constant of order unity.
Simple theoretical arguments \cite{FS97} suggest that this part of the distribution of the 
Wigner delay times is the same as in RMT considerations. This is in contrast with the
large delay times (see Eq.~(\ref{difftausmall})) where RMT considerations lead to a
power law decay (\ref{tauballlong}).

Since $\tau=\sum_{i=1}^M\tau_q$, we expect the behavior of the distribution of proper 
delay times ${\cal P}(\tau_q)$ to be similar to ${\cal P}(\tau)$ for large values of 
the arguments (for $\tau\gg1$ we have $\tau\sim \tau_q^{\rm max}$). 
The above predictions were verified numerically for a $2d$ diffusive system \cite{KOG03},
while unpublished results \cite{WMK05} show that the same law applies for a $3d$ diffusive
system as well.
We point out here that the asymptotic behavior ${\cal P}(\tau)\sim 1/\tau^{3/2}$ 
emerges also for chaotic/ballistic systems where the assumption of uniformly distributed 
$\delta E_n$ is the only crucial ingredient (see for example \cite{FS97}). 

%------------------------
%------------------------

%----------------------------------------------------------------------------------
\subsection{\bf Localized Regime}
\label{subsec:locdel}

In a serious of recent works \cite{TC99,OKG00,HLFPE99,JVK89} 
it was found that for 1D systems with $M=1$ and weak disorder the delay time distribution 
is
\begin{equation}
\label{delloc}
{\cal P}(\tau_W)= \frac {\xi}{v\tau_W^2} \exp(-\xi/v\tau_W),
\end{equation}
where $\xi$ is the localization length and $v= |\partial E/ \partial k|$ is the 
group velocity. (\ref{delloc}) 
takes its maximum value at $\tau_W^{max} = 0.5 \xi/v$, indicating that the most probable
trajectory that a particle travels (forth and back) before it scatters outside
the sample is the mean free path $l_{\rm mean} = \xi/4$. As $\tau_W\rightarrow
\infty$, ${\cal P }(\tau_W)$ shows a long time tail which goes as $2 \tau_W^{max}/
\tau_W^2$. This leads to a logarithmic divergence of the average value of $\tau_W$,
indicating the possibility of the particle traversing the infinite sample
before being totally reflected. As was indicated in \cite{OKG00,HLFPE99,JVK89}
this is another manifestation of the fact that in the
localized regime the conductance shows log-normal distribution due to the
presence of Azbel resonances. 

%------------------------
\begin{figure}
\begin{center}
\epsfxsize1\textwidth%
\includegraphics[width=10cm]{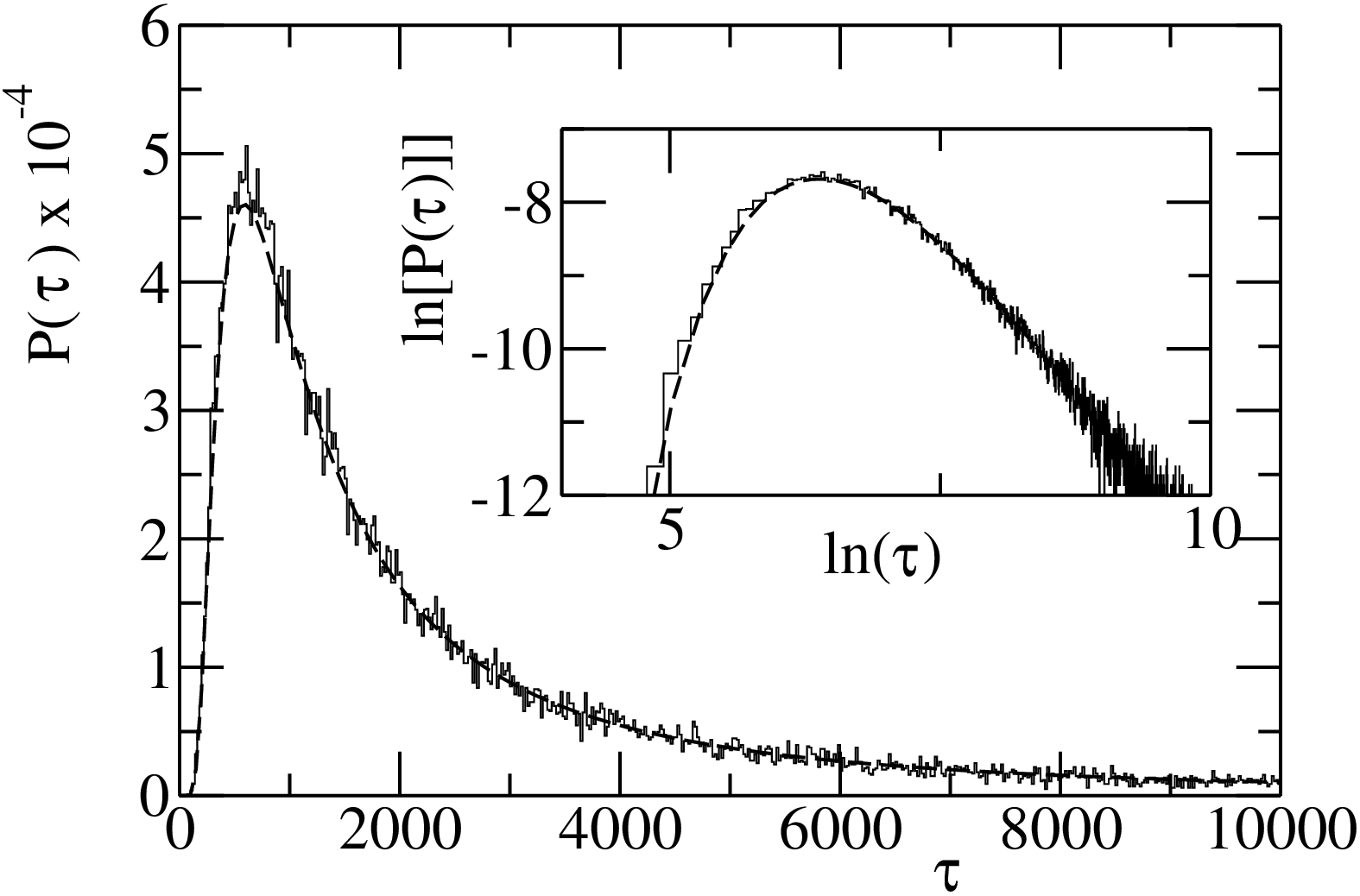}
%\epsfbox{fig10.ps}
\caption {\label{locdel}
Distribution of the delay times ${\cal P}(\tau_W)$ for the 1D Anderson model with on-site
potential, uniformly distributed between $[-0.1 ;0.1]$ and wavenumber $k=\sqrt
\pi$.  The dashed line corresponds to (\ref{delloc}). In the inset we present
the same data in a $\log-\log$ plot.
The figure is taken from \cite{OKG00}.}
\end{center}
\end{figure}   
%------------------------

We note that although the distribution for small delay times depends on disorder strength 
and possibly on number of channels $M$, the long time tail is universal. As a matter of
fact one can understand the long time power law behavior by employing the argument
leading to Eq.~(\ref{argtau}). Indeed by substituting the resonance distribution in the
localized regime (\ref{locres1}) we get again ${\cal P}(\tau_W)\propto 1/\tau_W^2$ independent
of the number of channels $M$. The validity of this calculation was checked recently \cite{WMK05}
for the $3d$ Anderson model in the localized regime and for $M\gg 1$ channels.

The above expression Eq.~(\ref{delloc}) does not contain the length of the chain, indicating 
that this intermediate asymptotics of the delay-time distribution, is related to the resonance 
width distribution, which is dominated by the electron escape rate from the resonant state into 
the nearest reservoir and for $L\rightarrow\infty$ is exact for any delay time. However, the 
finite length $L$ determines a cutoff $\tau_W\sim \exp^{L/\xi}$ for this universal 
behavior, and for larger delay times we find \cite{HLFPE99}
\begin{equation}
\label{delloc2}
{\cal P}(\tau_W) \propto \exp(-L/\xi) \tau_W^{-(1+{\xi\over L}\ln\tau_W)},\quad
\tau_W>\exp^{L/\xi}.
\end{equation}

%----------------------------------------------------------------------------------
\subsection{\bf Criticality }
\label{subsec:critdel}

Recently, an intensive activity to understand ${\cal P} (\tau_W)$ for systems at critical 
conditions was undertaken in \cite{KW02,F03,OF05,MK05}. The activity was mainly concentrated
in the simplest scattering set up of one open channel attached to the system of linear size 
$L$. As a result, an anomalous scaling, of inverse moments of $\tau_W$ with the system size 
$L$ was reported \cite{OF05,MK05} and specific predictions linking the scaling exponents and 
the multi-fractal properties of eigenfunctions of the corresponding closed system were made.
Specifically it was found that the inverse moments of Wigner delay times $\langle \tau_W^{-q} 
\rangle$ scale as 
\begin{equation}
\label{MIT}
\langle \tau_W^{-q} \rangle \propto L^{-f(q)},\quad f(q)= q D_{q+1}
\end{equation}
where $\langle . \rangle$ stands for an ensemble average. However it was found that this relation 
is extremely fragile \cite{MK05}. Namely, it holds for channels attached to a {\it typical} position 
inside the sample. This excludes the standard scattering set up where the channel is attached to 
the edge of the sample.

\begin{figure}
\begin{center}
\epsfxsize1\textwidth
\includegraphics[width=12cm]{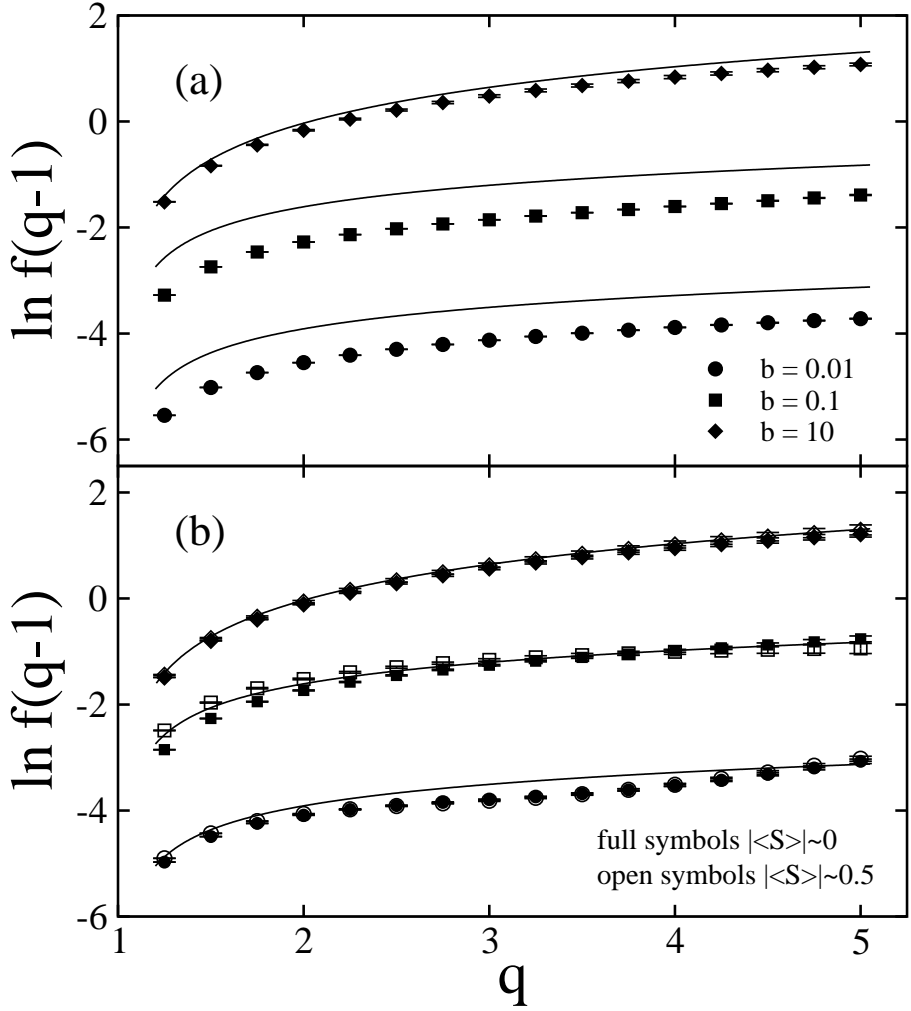}
%\epsfbox{fig11.eps}
\caption {\label{critdel} Scaling properties of inverse moments of delay times for the PBRM at criticality
\cite{MK05}. We report $\ln f(q-1)$ as a function of $q$ for a channel attached (a) to the boundary and (b) 
to the bulk of the sample when $|\langle S\rangle|\sim 0$ (full symbols). In (b) we also show $\ln f(q-1)$
for $|\langle S\rangle|\sim 0.5$ (open symbols). The curves are the theoretical prediction of Eq.
(\ref{MIT}). The figure is taken from \cite{MK05}. 
}
\end{center}
\end{figure}

In Fig.~\ref{critdel} we summarize the results of the investigation undertaken in \cite{MK05} 
where the analysis was performed for the Power Banded Random Matrix (PBRM) model, whose elements 
are independent random variables $H_{ij}$ with the variance decreasing in a power-law fashion: 
$\left<(H_{ij})^2\right>= [1+(|i-j|/b)^{2\alpha}]^{-1}$. For $\alpha=1$ this model shows
critical behavior and the fractal dimensions $D_q$ of the eigenfunctions depends on the parameter
$b$ and can be calculated analytically \cite{M00}. In Fig.~\ref{critdel}a we report the results 
for the case with a channel attached to the boundary. We see that the numerical data deviates 
from the theoretical predictions for any value of $b$. Instead, the agreement is very good for 
the case where the channel is attached to the bulk of the sample (see Fig.~\ref{critdel}b). In 
the latter case the channel is attached to a {\it representative} position in the sample. These 
results will provide a new method for evaluating the fractal dimensions $D_q$ in 
microwave and light wave experiments where $\tau_W$ can be extracted even in the presence of weak 
absorption. At the same time the appearance of the anomalous scaling of $\langle \tau_W^{-q}\rangle$ 
can be used as a criterion for detecting MIT.

%------------------------
%------------------------

%=======================================================================================
\section{\bf Quasi-periodic systems at criticality}
\label{sec:quasi}

Periodic and random media cover only the two extremes of the rich spectrum of complex systems.
Quasi-periodic systems \cite{AA80,SBGC84,MBCJB85,FO90,GKST94,GKP95,KS98,SOKG00,GONPSCW05} form 
an intermediate regime and have fascinating properties. In these deterministic non-periodic 
structures translational order is absent. In their one-dimensional tight-binding formulation 
they are described mathematically by the following Hamiltonian
\begin{equation}
\label{quasi}
\psi_{n+1}+\psi_{n-1}+W_n\psi_n=E\psi_n
\end{equation}
where $W_n$ is given by some quasiperiodic sequence. Among the most well studied 
representatives of this class are the Harper model \cite{AA80,GKP95,KS98,SOKG00} and Fibonacci 
quasi-crystals \cite{SBGC84,MBCJB85,FO90,GKST94,GKP95,SOKG00,GONPSCW05}. These two systems have 
been the subject of an extensive theoretical and experimental effort in the last twenty years.

The Harper model is described by the tight-binding Hamiltonian (\ref{quasi}) with on-site 
potential given by $W_n=\lambda \cos (2\pi\phi n )$. This system effectively describes 
a particle in a two-dimensional periodic potential in a uniform magnetic field with $\phi=
a^2eB/hc$ being the number of flux quanta in a unit cell of area $a^2$. When $\phi$ is an 
irrational number, the period of the effective potential $W_n$ is incommensurate with the 
lattice period. The states of the corresponding closed system are extended when $\lambda<2$, 
and the spectrum consists of bands (ballistic regime). For $\lambda>2$ the spectrum is point-
like and all states are exponentially localized (localized regime). The most interesting case 
corresponds to $\lambda=2$ of the MIT. At this point, the spectrum is a  Cantor set with fractal 
(box-counting) dimension $D_0^E\leq 0.5$ \cite{AA80,GKP95,SOKG00}. The spectral properties of the Harper model
were recently investigated in microwave experiments \cite{KS98}. Similar theoretical attention
was given also to the study of eigenfunctions  \cite{AA80,GKP95} which show self-similar 
fluctuations on all scales.

The Fibonacci binary quasi-crystal attracted a lot of interest as well. Here the potential
$W_n$ only takes the two values $+W$ and $-W$ arranged in a Fibonacci 
sequence \cite{SBGC84}. It was shown \cite{SBGC84,FO90,GKST94} that the spectrum is a Cantor 
set with zero Lebesgue- measure for all $W>0$. The first experimental realization of 
Fibonacci super-lattices was reported in \cite{MBCJB85} while their optical analogues were 
realized in \cite{GKST94,GONPSCW05}.

In \cite{SOKG00} we had presented consequences of the fractal nature of the spectrum in open 
quasi-periodic systems. We considered open systems with one channel (the simplest possible
scattering problem) and reported the appearance of a new type of resonances width and delay 
time statistics. These distributions show inverse power law behavior dictated by the fractal 
dimension $D_0^E$ of the spectrum. Specifically, it was found that the probability distributions 
of resonance widths ${\cal P} (\Gamma )$, and of delay times ${\cal P} (\tau )$ when generated 
over {\it different energies}, behave as
\begin{eqnarray}
\label{quasilaw}
{\cal P}(\Gamma)&=&\Gamma^{-\alpha}\,;\,\,\,\alpha=1+D_0^E\nonumber\\
{\cal P}(\tau_W)&=&\tau_W^{-\gamma}\,;\,\,\,\gamma=2-D_0^E\,\,.
\end{eqnarray}
Notice that for $D_0^E=0$ we recover the results (\ref{locres1},\ref{delloc}) of the point-like 
spectrum ($D_0^E=0$) corresponding to a localized system.

The connection between the exponents $\alpha, \gamma$ and the fractal dimension $D_0^E$ of the 
closed system calls for an argument for its explanation. The following heuristic argument
\cite{OKG00}, similar in spirit to the one used in Section \ref{subsec:balres}, provides some 
understanding of the power laws (\ref{quasilaw}). We consider successive rational approximants 
$\phi_i= p_i/q_i$ of the continued fraction expansion of $\phi$. On a length scale $q_i$ 
the periodicity of the potential is not manifest and we may assume that the variance of a wave 
packet spreads as $var(t)\sim t^{2D_0^E}$ \cite{GKP95}. We attach the lead at the end of the 
segment $q_i$ which results in broadening the energy levels by a width $\Gamma$. The maximum 
time needed for a particle to recognize the existence of the leads, is $\tau_{q_i}\sim q_i^{
1/D_0^E}$. The latter is related to the minimum level width $\Gamma_{q_i} \sim 1/\tau_{q_i}$.
The number of states living in the interval is $\sim q_i$ and thus determines the number of 
states with resonance widths $\Gamma > 1/ \tau_{q_i}$. Thus ${\cal P}_{int} (\Gamma_{q_i})\sim 
q_i \sim \Gamma^{-D_0^E}$. By repeating the same argument for higher approximants $\phi_{i+1}
=p_{i+1}/q_{i+1}$ we conclude that ${\cal P} (\Gamma)\sim \Gamma^{-(1+ D_0^E)}$, in agreement 
with (\ref{quasilaw}). Furthermore, use of Eq.~(\ref{argtau}) of Section \ref{sec:delayball} 
results in the distribution (\ref{quasilaw}) for the Wigner delay times. The numerical
results (see Figs. 12 and 13) obtained for the Harper and the Fibonacci models in \cite{OKG00}, 
verify the validity of the above arguments. Nevertheless a rigorous mathematical proof is still 
lacking.

%------------------------
\begin{figure}
\begin{center}
\epsfxsize1\textwidth%
\includegraphics[width=10cm]{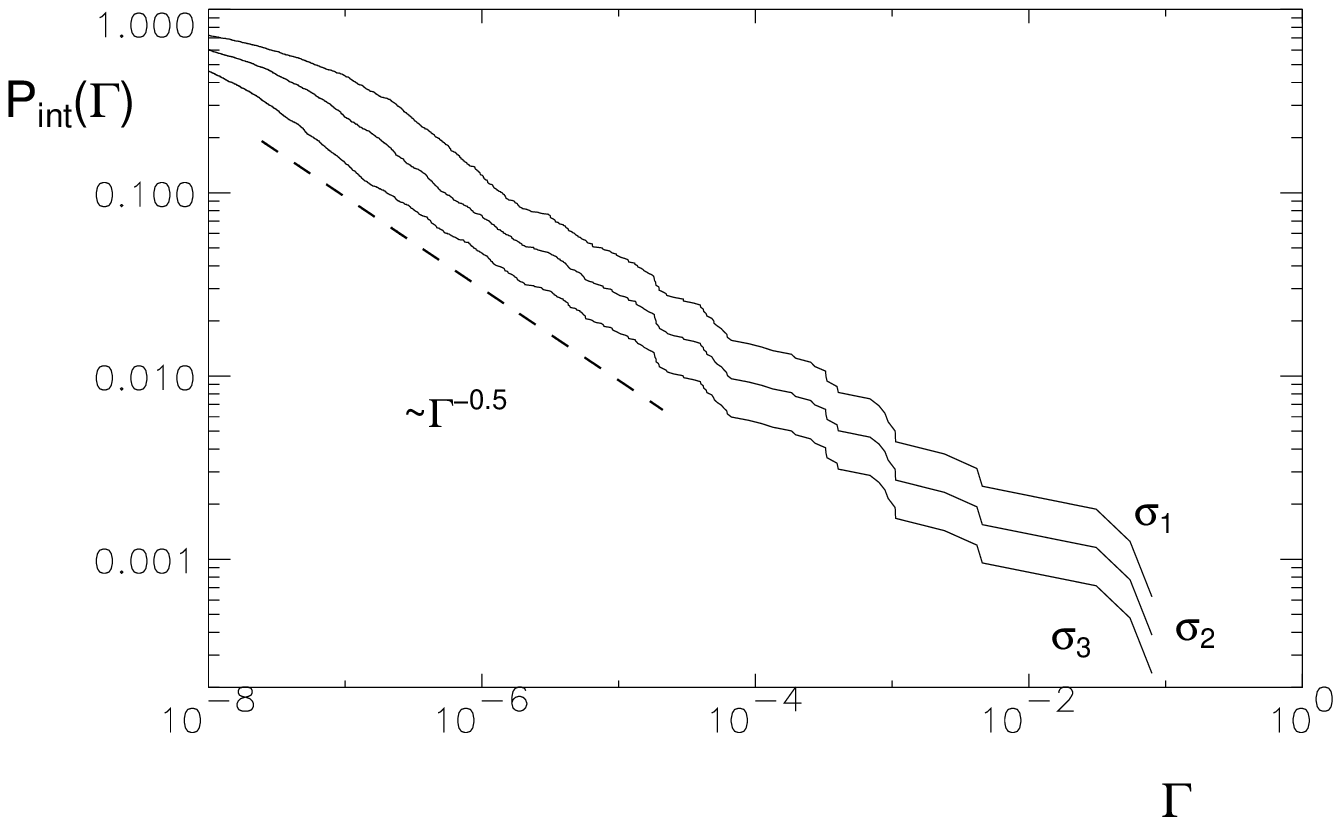}
\includegraphics[width=10cm]{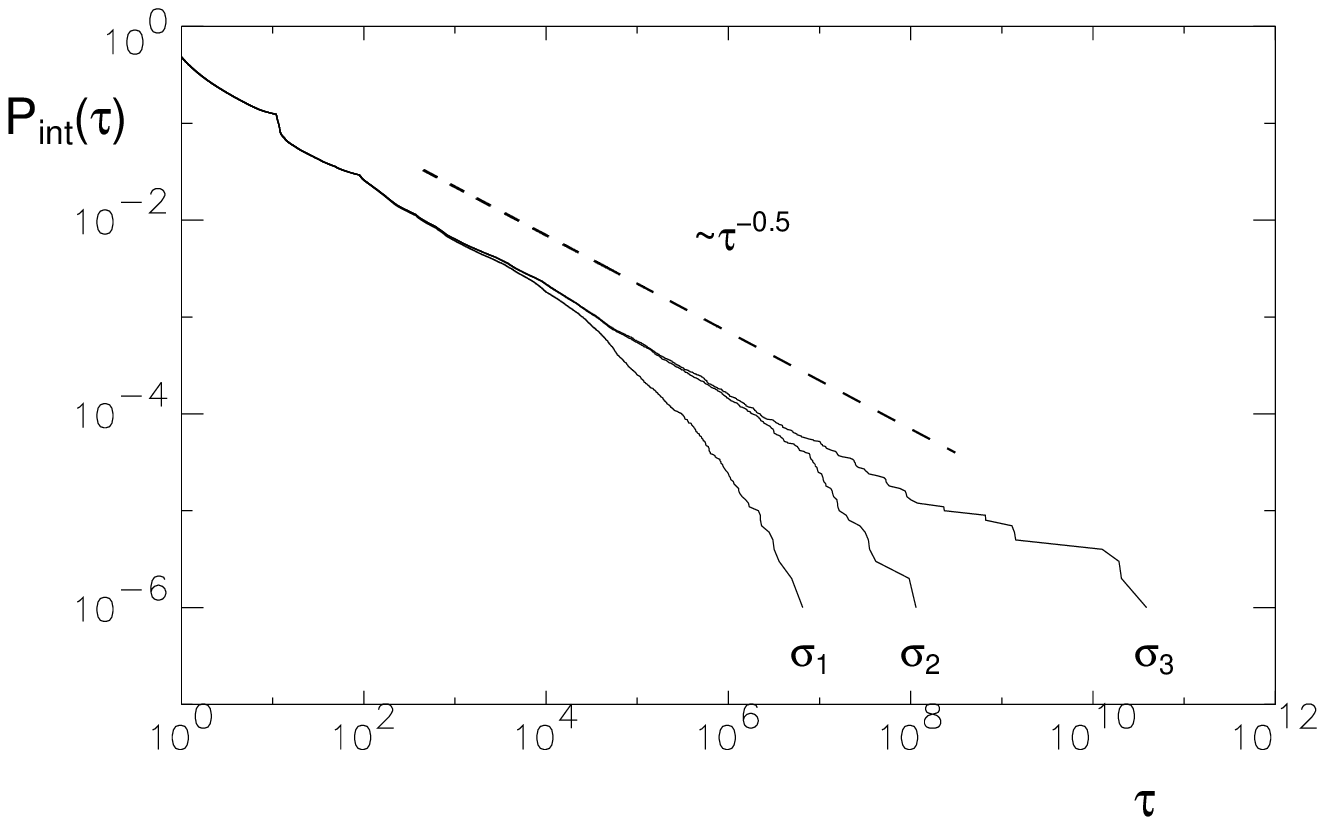}
%\epsfbox{fig12a.ps}
%\epsfbox{fig12b.ps}
\caption {\label{quasigamma}
(a)${\cal P}_{int}(\Gamma)$ of the Harper model $(\lambda=2)$
for three approximants of $\sigma_G$, $\sigma_1=\frac{987}{1597};
\sigma_2=\frac{1597}{2584};$ and $\sigma_3=\frac{2584}{4181}$.
An inverse power law $P_{int}(\Gamma) \sim \Gamma^{1-\alpha}$ is evident.
A least squares fit yields $\alpha
\approx 1.5$ in accordance with $D_0^E\simeq 0.5$ and Eqn.~(\ref{quasilaw}).
As is seen the lower cutoff of the scaling region decreases for higher
approximants.(b) ${\cal P}_{int}(\tau)$ of the Harper model $(\lambda=2)$ for three
approximants of the golden mean $\sigma_1=\frac{233}{377};
\sigma_2=\frac{987}{1597};$ and $\sigma_3=\frac{832040}{1346269}$.
An inverse power law $P_{int}(\tau) \sim \tau^{1-\gamma}$ is evident.
A least squares fit yields $\gamma\approx 1.5$ in accordance with
$D_0^E\simeq 0.5$ and Eq.~(\ref{quasilaw}). As is seen the upper
cutoff of the scaling region increases for higher approximants.
The figure is taken from \cite{SOKG00}. 
}
\end{center}
\end{figure}   
%------------------------

%------------------------
\begin{figure}
\begin{center}
\epsfxsize1\textwidth%
\includegraphics[width=10cm]{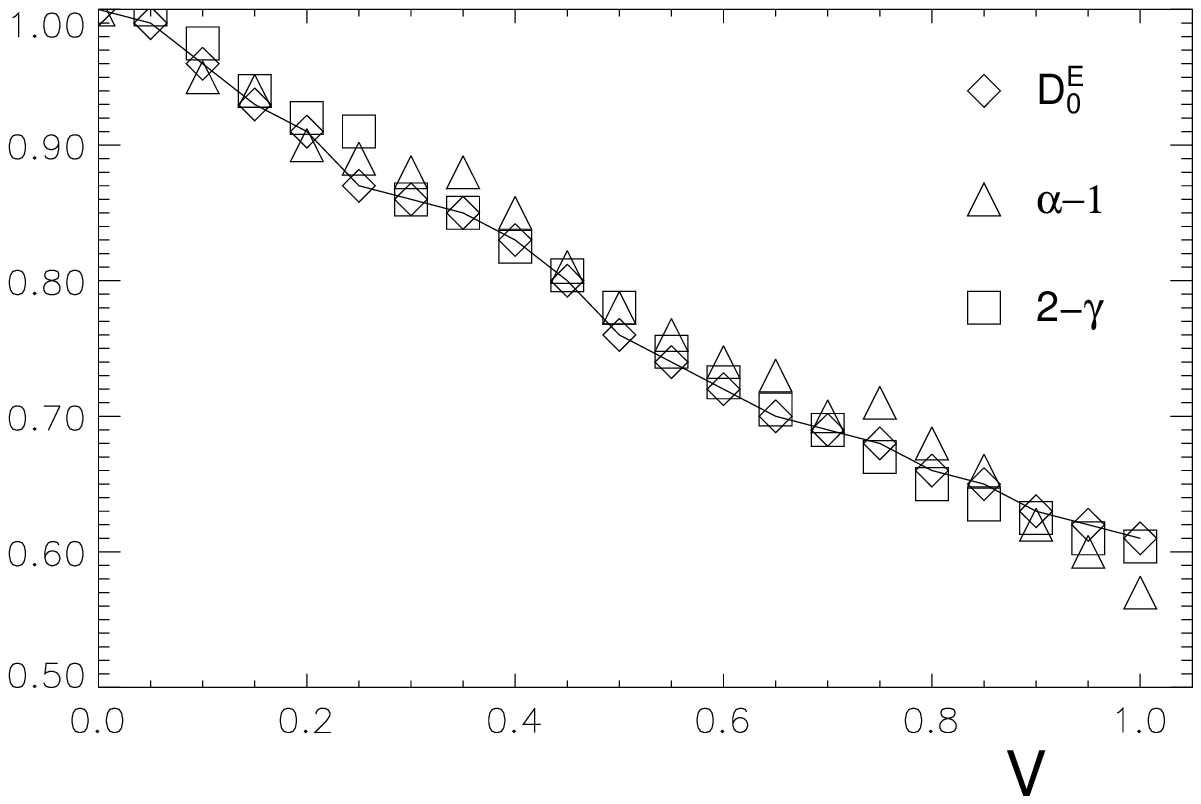}
%\epsfbox{fig13.ps}
\caption {\label{quasitau}
Power law exponents $\alpha,\gamma$ (plotted as $\alpha-1$ and $2-\gamma$)
of the resonance widths and of the delay time
distributions, respectively, as a function of the potential strength $W$ for
the Fibonacci model. We also plot the fractal dimension $D_0^E$ of the
spectrum (the solid line is to guide the eye). Our numerical data show
that $\alpha$ and $\gamma$ are related to the Hausdorff dimension $D_0^E$
according to Eqs.~(\ref{quasilaw}).
The figure is taken from \cite{SOKG00}. 
}
\end{center}
\end{figure}   
%------------------------

%=======================================================================================
\section{\bf Conclusions}
\label{sec:conclusions}

In this contribution we summarized the recent activity on the statistical properties of 
resonances and delay times of random/chaotic systems and analyze the deviations from 
the universal RMT predictions due to effects related to Anderson localization, diffusion 
and criticality. We have found that the tails of the resonance width distribution ${\cal P}
(\Gamma)$ reflects the nature of the {\it dynamics} associated with the corresponding 
closed system as it is defined by the second moment of a spreading wavepacket (ballistic, 
diffusive, critical, or localized). Instead, the origin, corresponding to small resonance 
widths, is dictated by anomalously localized states. Moreover, we have found that in the 
diffusive regime ${\cal P}(\Gamma)$ is affected by the channel "configuration" (position 
and relative number) as well, in contrast to the localized regime. At MIT the resonance
width distribution is universal and can be used to formulate a scaling theory of localization.

Localization phenomena affect also the delay times leaving their traces to the the 
distributions ${\cal P}(\tau)$. For scattering systems attached to one open channel, we 
have a very good quantitative understanding of ${\cal P}(\tau)$ for all regimes. A general expression
connects this distribution with the distribution of wavefunction intensities, the latter
being well studied during the last years. This connection allow us to use the experimentally 
accessible delay times in order to probe properties of wavefunctions, like multifractality, 
which are not easily measured. For many open channels, ample of numerical data supported 
by theoretical arguments allow us to estimate the shape of the distribution of delay times 
and get a qualitative understanding of the traces of localization. Nevertheless it remains 
a challenge to get some quantitative expressions as well.

The last section of this review is devoted to the $1d$ quasi-periodic systems 
at criticality. The corresponding closed systems show fascinating properties like spectral 
and wavefunctions fractality. We reviewed the traces of spectral fractality to the distributions 
of the resonance widths ${\cal P} (\Gamma)$, and of delay times ${\cal P} (\tau)$. Based
on numerical results and theoretical arguments it is shown that both quantities decay 
algebraically with powers which are related to the fractal (box counting) dimension $D_0^E$ 
of the spectrum of the closed system. 

%=======================================================================================
\section{\bf Acknowledgments}

We acknowledge many useful discussions with Y. Fyodorov, T. Geisel, A. Mendez-Bermudez,  
A. Ossipov, D. Savin, H. Schanz, U. Smilansky and M. Weiss. This work was supported by 
a grant from the GIF, the German-Israeli Foundation for Scientific Research and Development.

%-----------------------------------------------------------------------------

\end{document}